\documentclass[paper]{JHEP}
\usepackage{amsmath}
\usepackage{cite}

\def\cO#1{{\cal{O}}\left(#1\right)}

\def\cP{{\cal{P}}}
\def\cS{{\cal{S}}}
\def\MSbar{\overline{\mbox{\scriptsize MS}}}
\def\al{\alpha}
\def\be{\beta}
\def\Om{\Omega}
\def\ca{C_A}
\def\cf{C_F}
\def\half{\mbox{\small $\frac{1}{2}$}}
\def\cM{{\cal{M}}}

\def\citd#1#2{\cite{#1,#2}}

\def\EEC{{\rm EEC}}

\def\ae{\alpha_{\mbox{\scriptsize eff}}}

\def\lqcd{\Lambda_{\mbox{\scriptsize QCD}}}
 \def\cR{{\cal{R}}}
\def\NP{{\mbox{\scriptsize NP}}}
\def\PT{{\mbox{\scriptsize PT}}}

\def\la{\mathrel{\mathpalette\fun <}}
\def\ga{\mathrel{\mathpalette\fun >}}
\def\fun#1#2{\lower3.6pt\vbox{\baselineskip0pt\lineskip.9pt
  \ialign{$\mathsurround=0pt#1\hfil##\hfil$\crcr#2\crcr\sim\crcr}}}

\def\eps{\epsilon}
\def\de{\delta}
\def\as{\alpha_{\mbox{\scriptsize s}}}
\def\merge{\mbox{\scriptsize merge}}

\newcommand\epjc[3]{  
                        {{\it Eur. Phys. J. }{\bf C #1} (#2) #3}}

\title{On the universality of the Milan factor for $1/Q$ power
    corrections to jet shapes\thanks{ Research supported in part by
      MURST, Italy.}}
\author{Yu.L. \ Dokshitzer\thanks{On leave from St. Petersburg Nuclear
    Institute, Gatchina, St. Petersburg 188350, Russia}, A.\ Lucenti,
  G.\ Marchesini and G.P.\ Salam \\
  Dipartimento di Fisica, Universit\`a di Milano \\
  and INFN, Sezione di Milano, Italy \\}

\abstract{We perform the two-loop analysis of the $1/Q$ power
  corrections to jet-shape variables.  This step is necessary for
  producing reliable theoretical predictions for the relative
  magnitudes of genuine confinement effects.  We show that the
  rescaling factor recently derived for the thrust case (the Milan
  factor) remains the same for the class of observables which includes
  the $C$-parameter, invariant jet masses, jet broadening and the
  energy-energy correlation measure.  We list the expressions which
  should be used for extracting $1/Q$ power effects in jet shapes.
  We also envisage large non-perturbative effects, characterised by
  {\em fractional}\/ powers of $Q$, in certain semi-inclusive
  observables such as the height of the spectrum of Drell-Yan lepton
  pairs with invariant mass $Q$ and transverse momentum $Q_t\!=\!0$,
  and back-to-back energy flows in $e^+e^-$ and DIS.  }

\keywords{QCD, NLO Computations, Jets, LEP HERA and SLC Physics}

\preprint{IFUM-601-FT\\
  hep-ph/9802381 \\
  February 1998}

\begin{document}

\section{Introduction}

There is a set of jet-shape observables that contain  
$1/Q$ power correction terms~\cite{Web94,DW,StKoZakh}.
These non-perturbative contributions originate from soft-gluon
emission with energies of the order of the confinement scale.

The standard technology for determining these power terms relies on
the introduction of a fake gluon mass.  In a more systematic approach,
it is the variable $m^2$, entering in the dispersive representation
for the running QCD coupling, that plays the r\^ole of the power
correction trigger~\cite{BB95b,BBB95,DMW}.  Within this approach the
magnitude of the power contribution is expressed in terms of a
standard dimensional integral of the non-perturbative part of the QCD
coupling (``non-perturbative effective coupling modification''),
multiplied by an observable-dependent (easily calculable) coefficient.

In both cases the gluon decays, essential for building up the running
coupling, were treated inclusively.  This is known as the naive
approximation.  It is based on replacing the actual contribution of
the final-state partons by that of their massive parent gluon. This
procedure is intrinsically ambiguous since there is no
unique prescription for including finite-$m^2$ effects into the
definition of the shape variable.  Moreover, Nason and
Seymour~\cite{NS} questioned the application of the inclusive
treatment of gluon decays.  They pointed out that jet shapes are not
inclusive observables, and therefore that the configuration of
offspring partons in the gluon decay may affect the value of the power
term at next-to-leading level in $\as$, which a priori is no longer a
small parameter since the characteristic momentum scale is low.  Both
these problems can be resolved only at the two-loop level.

For the thrust case the problem was addressed recently
in~\cite{DLMSthrust}.  The two-loop analysis led to a modification of
the naive prediction for the coefficient of the $1/Q$ term by a
perturbatively calculable $C_A$- and $n_f$-dependent numerical factor,
which has since been named ``the Milan factor''~\cite{BryanFrascati}.
A priori one might expect the Milan factors to differ for different
observables.

In this paper we consider, in addition to thrust, the following
infrared- and collinear-safe observables: the $C$-parameter, invariant
jet masses and jet broadening.  In near-to-two-jet kinematics (soft
limit) these observables are linear in secondary particle momenta, and
therefore produce $1/Q$ terms.

We compute Milan factors for these observables, which rescale the
naive predictions.  We show that given an appropriate, consistent,
definition of the naive approximation, the observables belonging to
the same $1/Q$ class possess a universal Milan factor, $\cM\simeq
1.8$.

We address the problem of constructing renormalon-free answers that
combine the truncated perturbative series with the power-behaving
non-perturbative contributions.  We choose to quantify the magnitudes
of the non-perturbative contributions by means of finite-momentum
integrals of the running QCD coupling $\as$ rather than in terms of
the moments of its dispersive partner, the so-called ``effective
coupling'' $\ae$~\cite{DMW}, as suggested in~\cite{DW}.  The
corresponding numerical ``translation factor'' $2/\pi$ happens to
practically compensate the effect of the Milan factor. As a result, we
expect the phenomenology of $1/Q$ power contributions to jet shapes to
be affected only slightly.

The same Milan factor applies to the $1/Q$ power correction to the
energy-energy correlation measure in $e^+e^-$ annihilation (EEC)
outside the region of back-to-back kinematics.  A qualitative analysis
of $\EEC$ in the special back-to-back kinematical region shows that
one can expect large non-perturbative corrections characterised by 
fractional powers $Q^{-\gamma}$, with $\gamma<1$.  This results from an
interplay between non-perturbative and perturbative effects, which is
typical for semi-inclusive observables.  A fractional power should also
be present in the distribution of heavy lepton pairs with small
transverse momenta (the Drell-Yan process).

The paper in organised as follows: in section 2 we define and describe
the observables in the soft radiation limit; section 3 is devoted to
the all-order resummation programme; in sections 4 and 5 we extract
the leading $1/Q$ power corrections to the radiator and to the
corresponding distributions; section 6 is devoted to the merging of
the perturbative and non-perturbative contributions in a
renormalon-free manner; we summarise the results and present
conclusions in section 7.

\section{Kinematics and observables}\label{s:KO}

Here we consider the kinematics of multi-particle ensembles consisting
of a primary quark and antiquark with 4-momenta $p$ and $\bar p$, and
$m$ secondary partons $k_i$.  Hereafter we shall consider the phase
space region of soft secondary partons, soft gluons and their decay
products, so that the primary $p$ and $\bar{p}$ belong to the opposite
hemispheres.

We introduce the standard light-cone Sudakov decomposition 
in terms of two vectors chosen in the direction of the thrust axis:
$$
 P^\mu+\bar{P}^\mu=Q^\mu;\> P^2=\bar{P}^2=0;\> 2(P\bar{P})=Q^2 \equiv 1\,.
$$
The parton momenta are:
\begin{eqnarray*}
        p&=& \alpha_p \bar{P} + \beta_{p}P + p_t \>, \\
 \bar{p}&=& \alpha_{\bar{p}} \bar{P} + \beta_{\bar{p}}P + \bar{p}_t \>; \\
      k_i &=&  \alpha_i \bar{P} + \beta_{i}P + k_{ti} \>.
\end{eqnarray*}
For a massless particle $q$, we have $\alpha_q\beta_q=q_{t}^2$.
In the soft approximation, for the primary quarks we have
\begin{equation}\label{softpp}
 (1-\alpha_{\bar{p}}),\> (1-\beta_p) \> \ll 1\>, \qquad   
 \beta_{\bar{p}}\simeq \bar{p}_{t}^2, \>  
  \alpha_{{p}}\simeq {p}_{t}^2 \> \ll\> 1\,,
\end{equation}
while all the secondary-parton momentum components are small.
The differences $1-\alpha_{\bar{p}}$ and $1-\beta_p$
are {\em linear}\/ in gluon momenta and will be taken care of;   
the components  $\beta_{\bar{p}}$ and $\alpha_{{p}}$ are 
{\em  quadratic}\/ and will be neglected in what follows. 
 
The jet shapes defined with the use of the thrust axis are $T$, the
invariant squared mass of all the particles in one hemisphere (``jet
mass'' $M^2$) and the sum of the moduli of the momentum components
transverse to the thrust axis (the jet broadening $B$~\cite{Bdefref}).

In terms of Sudakov variables they read:
\begin{eqnarray}
\label{Tdef}
1-T &=& \sum_{i=1}^N \min(\alpha_i,\beta_i)+\alpha_p+\beta_{\bar p}\>, \\
\label{M2def}
  M^2_R 
        &=& \left(\alpha_p + \sum_{i\in R} \alpha_i\right)
          \left(1 - \beta_{\bar p} - \sum_{i\in L} \beta_i\right) , \\
\label{Bdef}
 2B_R   &=& 
  \sum_{i\in R} \left|\vec{k}_{t i}\right| + 
  \left|\sum_{i\in R} \vec{k}_{t i} \right| .
\end{eqnarray}
Here we have defined the right hemisphere as containing the quark $p$,
so that for a particle in the right hemisphere, $\alpha < \beta$
(correspondingly, partons with $\beta<\alpha$ contribute to $M^2_L$,
$B_L$).  We have also used the fact that the total transverse momentum
of particles in each hemisphere is zero, which is a property of the
thrust axis.

There are two types of invariant mass and broadening distributions
under consideration. One can discuss the total squared mass, the total
broadening, etc., which are characteristics of the whole event:
$$
M_T^2 \>=\> M^2_R+M^2_L\>, \qquad  B_T \>=\> B_R+B_L\>. 
$$
Alternatively, one can study the {\em heavy-jet}\/ mass, the
{\em wide-jet}\/ broadening:
$$
  M^2_H \>=\> \max\left\{ M_R^2, M_L^2\right\}, \qquad
  B_W \>=\> \max\left\{ B_R, B_L\right\}.
$$
The all-order perturbative analysis has been carried out both for the
total and for single-jet (``heavy'', ``wide'')
characteristics~\cite{CTTW,CTW}. 
As we shall see later, the non-perturbative effects in the
single-jet characteristics, in general, amount to {\em half}\/ of 
those in the {\em total}\/ event observables.

The $C$-parameter and the energy-energy correlation function (EEC) are
defined as
\begin{eqnarray}
\label{Cdef}
  C &=& \frac32 \sum_{ab} p_a p_b \sin^2 \theta_{ab} \>, \\
\label{EECdef}
  \EEC(\chi) &=& \sum_{ab} E_a E_b\; \delta(\cos\chi  -\cos\theta_{ab})\>,
\end{eqnarray}
where $p_a$ and $E_a$ are the modulus of the 3-momentum and the energy
of parton $a$ (the quark, the antiquark or one of the secondary
partons); $\theta_{ab}$ is the angle between the two momenta. The sums
over $a,b$ run over all particles in an event.  (Notice that within
the definition of the $C$ parameter and EEC, each pair $a,b$ is
counted twice.)  In the massless particle approximation one has the
sum rule
\begin{equation}\label{sumrule}
  C = \frac32 \int_{-1}^1 d(\cos\chi) \; \EEC(\chi) \;\sin^2\chi\,. 
\end{equation}

Now we consider each of these observables in the soft limit.

\paragraph{Thrust.}
If only one gluon is emitted, say in the right hemisphere,
$\alpha_1<\beta_1<1$, then 
$$
 1-T = \alpha_1+\alpha_{p}+\beta_{\bar{p}}\;.
$$
Since the thrust axis is such that $k_{t1}^2=p_{t}^2$, we have
$$
 \alpha_1=\frac{p_{t}^2}{\beta_1}\;, \quad 
 \alpha_{p}=\frac{p_{t}^2}{1-\beta_1}\;, \quad
 \beta_{\bar{p}}=0\;.
$$
The differential thrust distribution takes the form~\cite{Farhi}
$$
(1\!-\!T) \frac{d\sigma}{\sigma\,dT} \>=\> 
 \frac{C_F\alpha_s}{\pi}\left( 4\ln\frac{1}{1\!-\!T}-3+\cO{1\!-\!T}\right).
$$
Here the logarithmic piece comes from the soft limit neglecting the
quark contribution:
$\alpha_p=p_t^2\ll \alpha_1=p_t^2/\beta_1$, so that $1-T\simeq \alpha_1$.
The constant term accounts for  gluon and quark contributions 
in the region of relatively hard gluons: $\beta_1\sim \beta_p\sim 1$,
$\alpha_1\sim \alpha_p$.

In higher orders the soft gluon contributions exponentiate into the
double-logarithmic Sudakov form factor, while the effects of hard
gluons and of quark recoil remain subleading, being down by one power
of $\ln(1-T)$.  However, the quark contributions to thrust,
$\alpha_p+\beta_{\bar{p}}$, don't contribute at all to the $1/Q$
power-suppressed contribution, which is driven exclusively by soft
gluons.  Therefore hereafter we ignore these quark contributions to
(\ref{Tdef}), though we note that the perturbative part of the answer
takes them into proper account~\cite{CTTW}.

Defining 
$$
  A\>=\>  
\sum_{i\in R} \alpha_i\>, \quad  \bar{A}\>=\> \sum_{i\in L} \beta_i \>,
$$
and neglecting the quark contributions we have 
\begin{equation}
  \label{1mT}
  1-T \>=\> A+\bar{A}\>.
\end{equation}

\paragraph{Jet mass.}
It is common in the discussion of jet masses, in addition to
considering the sum of the squared masses of the two hemispheres,
$$
M_T^2  = M_R^2 + M_L^2 \,,
$$
to study the heavy jet mass,
$$
M_H^2 = \max\{M_R^2,M_L^2\}\,.
$$
To simplify the analysis we postpone the consideration of which
hemisphere is heavier and discuss the double-differential distribution
in $M_R^2$ and $M_L^2$. This will help avoid the considerable
confusion that arises in a direct $\as^2$ analysis of the
non-perturbative effects in $M_H^2$. Indeed, aiming at power effects,
we should treat the two hemispheres on an equal footing, since it is
the {\em perturbative}\/ radiation which determines which of the two jets is
heavier.

The quark contributions to the jet mass (\ref{M2def}) 
can be analysed and dropped for the study of the $1/Q$ power
corrections, as in the thrust case, so that we have
\begin{equation}\label{M2AB}
 M^2_R= A(1-\bar{A})\>,  \quad  M^2_L= \bar{A}(1-A)\>. 
\end{equation}
In the first order in $\as$ the relation between the jet masses and the 
thrust reads 
\begin{equation}\label{Tlinear}
 1-T \>\simeq\> M_R^2+M_L^2\>=\>M_T^2\>.
\end{equation}
We observe that at the $\alpha_s^2$ level this relation 
becomes non-linear. 
From \eqref{1mT} and \eqref{M2AB} one derives
\begin{equation}\label{Tnonlinear}
 (1-T)\frac{1+T}{2} \>=\>  M_R^2+M_L^2-\frac{(M^2_R-M^2_L)^2}{2}\>.
\end{equation}
This introduces corrections to (\ref{Tlinear}) of relative order
$M_{R,L}^2/Q^2$.  In the region of small jet masses these effects are
negligible.  At the same time, averaging over, say $M_L^2$ in the
$M_R^2$-distribution will produce a cross-talk correction at a
relative level of $\left\langle M_L^2/Q^2\right\rangle\sim
\alpha_s(Q^2)$.  As a result, the power term $1/Q$ in which we are
interested will also acquire a relative correction of order
$\alpha_s(Q^2)$.  This subleading perturbative rescaling is beyond the
accuracy that we consider.

\paragraph{Broadening.}
The distribution of the jet broadening, $B$, is different in two
respects from those of the other variables considered here. At the
perturbative level, the quark recoil can not be neglected. In the
first order, for example, the quark contribution to $B$ is simply
equal to that of the gluon. Hence, the factor of two in the definition
of $B$~\eqref{Bdef}.  In higher orders an account of quark recoil
becomes more complicated~\citd{CTW}{DLMSbroad}.
At the non-perturbative level, broadening has an extra collinear
enhancement, leading to a power correction proportional to $(\ln
Q)/Q$.  Broadening is the only observable currently under discussion
with such a type of leading power correction.  One can expect
analogous (i.e.\ $\ln Q$-enhanced) confinement effects in the $E_t$
distribution in hadron-hadron collisions, in the oblateness measure,
etc.

\paragraph{$C$-parameter.}
In the soft limit, that is in the linear approximation in gluon
energies, the relevant contributions to $C$ come from counting (twice)
the gluon-quark pairs,
\begin{equation}
   \label{Csoft0}
 C \>\simeq\> \frac34\cdot 2\left\{ 
  \sum_{i} p^0  k^0_i \, \sin^2\Theta_{ip} 
+ \sum_{i} \bar{p}^0  k^0_i \, \sin^2\Theta_{i\bar{p}} \right\} .
\end{equation}
The gluon-gluon contributions are quadratic in gluon energies and
therefore negligible.  The quark-antiquark contribution can also be
neglected since $\sin^2\Theta_{p\bar{p}}$ is proportional to the total
gluon transverse momentum squared, and thus is also quadratic in gluon
energy.  Setting $p^0=\bar{p}^0=\half$ and making use of the Sudakov
variables, $k^0_i=\half(\al_i+\be_i)$ and
$\sin^2\Theta_{ip}=\sin^2\Theta_{i\bar{p}}
=4\al_i\be_i/(\al_i+\be_i)^2$, we get
\begin{equation}
   \label{Csoft}
    C \>=\>  6 \sum_i \frac{\al_i\be_i}{\al_i+\be_i}\>. 
\end{equation}

\paragraph{Energy-Energy Correlation.}
The problem of power effects in energy-energy correlations has two
aspects. In the region of finite angles, $\chi\sim 1$, the leading
$1/Q$ power term originates from triggering a gluon with an energy 
of order of the confinement momentum scale. 
This contribution does not involve gluon resummation  
and can be obtained within a fixed order analysis.
It is straightforward to derive the EEC function 
in the soft approximation:
\begin{equation}
  \label{EECsoft}
  \EEC(\chi)\>=\> \frac2{\sin^3\chi} \sum_i k_{ti}\, 
\delta\left(1-\frac{\al_i}{k_{ti}\tan\frac{\chi}{2} }\right). 
\end{equation}
As we shall see later, the power correction explodes at $\sin\chi\to0$
where multi-parton resummation effects become essential.

\section{Observables and the resummation programme}
In what follows we concentrate on observables which require all-order
soft gluon resummation to be carried out at the two-loop level. We
introduce the label $V$ as a general name for the observables
$1\!-\!T$, $B$, $C$ and the double jet-mass, so as to treat them
simultaneously.  For the broadening we consider the observable
$B=B_T$. EEC away from the back-to-back region doesn't require
resummation, and will be addressed at the end.

\subsection{Observables}
We consider the $V$-distributions in the region $V\ll1$.
The observables $1\!-\!T$, $C$ and $B$ are given as a sum of 
contributions $v(k_i)$ from each emitted parton $k_i$.
They can be expressed in terms of the multi-parton emission 
distribution $d\sigma_n$ by the following relations
\begin{equation}
  \label{sBR}
\begin{split}
  \frac{d\sigma}{\sigma (d\ln V)} =&
  \frac{d}{d\ln V}\> I(V)\>,\\
  I(V)=& 
    \sum_m \int \frac{d\sigma_m}{\sigma} \>\> 
    \Theta\left(V -\sum_{i=1}^m v(k_i) \right) \,,
\end{split}
\end{equation} 
with $m$ the total number of secondary partons in the final state
(soft gluons or their decay products).  We have $V=1\!-\!T,\, C$ and
$2B$ (notice that $B$ is defined in \eqref{Bdef} with a factor $2$, as
a {\em half-sum}\/ of the moduli of the transverse momenta).

For small $V$, which corresponds to all the final-state partons 
having small transverse momenta, one can write $d\sigma_m$ as a 
product of two factors, 
$$
   \frac{d\sigma_m}{\sigma} \>=\> C(\as)\, dw_n\>.
$$
The first is an observable-dependent coefficient factor 
which is a function only of $\as(Q)$. 
The second is the observable-independent 
``evolutionary exponent'', which describes the 
production of $n$ small-$k_t$ gluons off the primary quark-antiquark
pair and their successive splitting into $m\ge n$ final partons.
Hence
\begin{equation}
  \label{universal}
  I(V) = C(\as(Q))\> \Sigma(V,\as) \,, \qquad 
  \Sigma(V,\as)=  \sum_n \int dw_n \>\> 
    \Theta\left(V -\sum_{i=1}^m v(k_i) \right).
\end{equation}
We note that $dw_n$ implicitly contains a sum over $m$ in the range
$n\le m\le 2n$.
The soft distribution $dw_n$ is normalised to unity: 
$\sum_n \int dw_n =1$, where the transverse momentum of each gluon is
integrated up to $Q$.
The independent emission distribution is valid only for small $k_{ti}$.
It simplifies the treatment but mistreats the non-logarithmic region
of large transverse momenta, $k_t\sim Q$, both in real and virtual
terms. 
This is compensated by the factor $C(\as)$ in \eqref{universal}.
For small $V$ values this factor is constant. 
The dependence on $\ln V$ is embodied into the $\Sigma$ factor.

The essential momentum scales in the coupling in $\Sigma$ range from
$V Q$ to $Q$. 
The perturbative treatment requires 
$V Q \gg \Lambda_{\mbox{\scriptsize QCD}}$. 
The power correction we are interested in originates from 
$v(k_i)Q\sim \lqcd$. 

The double-differential jet-mass distribution can be obtained via 
the distribution in $A,\bar{A}$ (see \eqref{M2AB}) as 
\begin{equation}
\frac{\sigma ^{-1}\>d\sigma}{dM_R^2 dM_L^2}\>=\> C(\as(Q))
\int dAd\bar{A}\> \frac{d^2\Sigma(A,\bar{A})}{dAd\bar{A}}\>
\delta(M_R^2-A(1-\bar{A}))\>\delta(M_L^2-\bar{A}(1-A))\,.
\end{equation}
The double-distribution $\Sigma(A,\bar A)$ is given by
\begin{equation}\label{SD}
\Sigma(A,\bar A)= \sum_n\int dw_n
\>\Theta(A-\sum_{i\in R}\al_i)
\>\Theta(\bar{A}-\sum_{i\in L}\beta_i)\>.
\end{equation}
We denote it as the $D$-distribution (with $D$ standing for {\em
  double}\/).

\subsection{Soft multi-parton ensembles}
To obtain the multi-parton soft-emission formula at the two-loop level 
it suffices to take into account a single splitting of each of the
gluons radiated off the primary $q\bar{q}$ pair. 
It has the form
\begin{multline}\label{dwn}
dw_n = \frac{1}{n!}\prod_{i=1}^n
\left\{
\frac{C_F}{\pi}\frac{d\al_i}{\al_i}\frac{d^2k_{ti}}{\pi k_{ti}^2}
[\as(0)+4\pi\chi(k_{ti}^2)]
+4C_F\>d\Gamma_2(k_i,k_i')\left(\frac{\as}{4\pi}\right)^2
\frac 1{2!}M^2 \right\} \\
\exp\left\{
-\frac{C_F}{\pi}\int \frac{d\al}{\al}\frac{d^2k_{t}}{\pi k_{t}^2}
[\as(0)+4\pi\chi(k_{t}^2)]
-4C_F\int d\Gamma_2(k_1,k_2)\left(\frac{\as}{4\pi}\right)^2
\frac1{2!}M^2\right\} .
\end{multline}
This expression has the following structure. 
The first term describes the real emission of a soft massless gluon 
(the term with an ill-defined $\as(0)$)
and the two-loop virtual contribution to it 
(the renormalisation-scheme-dependent function $\chi(k^2_t)$). 
The last term on the first line describes the gluon splitting into
a $gg$ or $q\bar{q}$ pair. 
The corresponding matrix element $M^2$ is
given in~\cite{DLMSthrust}. 
Finally, the exponential factor on the second line stands for the
total virtual correction factor which ensures the normalisation of $dw_n$.
The expression \eqref{dwn} embodies the production of up to $m=2n$
secondary final partons.  

The two-parton phase space is
\begin{equation}
d\Gamma_2(k_1,k_2) 
\;=\; \prod_{i=1}^2\frac{d\al_i}{\al_i} \frac{d^2k_{ti}}{\pi} 
\;=\; \frac{d\al}{\al}\frac{d^2k_t}{\pi}\frac{d^2q_t}{\pi}\>
z(1-z)\>dz\,,
\end{equation}
where $\al=\al_1+\al_2$, $\vec{k}_t$ is total transverse 
momentum of the parent gluon, 
$\vec{q}_t$ is the relative ``transverse angle'' of the pair, and $z$
is the fraction $\al_1/\al$. 
We have  
\begin{equation}
  \label{kin}
\al_1=z\al\,,   \quad   \al_2=(1-z)\al\,,  
\quad \vec{k}_t=\vec{k}_{t1}+\vec{k}_{t2}\,,
\quad \vec{q}_t=\frac{\vec{k}_{t1}}{z}-\frac{\vec{k}_{t2}}{1-z}\,.
\end{equation}
We introduce the rapidity and the mass of the parent gluon by
\begin{equation}
  \label{eta}
m^2=z(1-z)q_t^2\,;
\qquad \al=\sqrt{k_t^2+m^2}e^{-\eta}\>,  
\qquad \be=\sqrt{k_t^2+m^2}e^{\eta}  \>.
\end{equation}
The right (left) hemisphere corresponds to $\eta>0$ ($\eta<0$).

In these terms the two-parton phase space reads
\begin{equation}
d\Gamma_2(k_1,k_2) 
\>=\> dm^2\>d\eta\>\frac{d^2k_{t}}{\pi} \>\>dz\>\frac{d\phi}{2\pi} \>,
\end{equation}
where $\phi$ is the angle between $\vec{k}_t$ and $\vec{q}_t$.

\subsection{$\Sigma$--distribution}
Taking advantage of the factorised structure of the multi-gluon matrix
element \eqref{dwn} we introduce the source function
$$
 u(k)=u(\al,\be) \>\equiv\> \exp(-\nu v(k))
$$
to write in terms of the Mellin integral transform 
$$
 \Theta\left(V-\sum_{i=1}^m v(k_i)\right) \>=\> 
\int \frac{d\nu}{2\pi i \, \nu} \>\> e^{\nu V}\> \prod_{i=1}^m u(k_i)\>.
$$
Then the $\Sigma$--distribution takes the form, for $V=1\!-\!T$,
$C$ and $2B$,
\begin{equation}
  \label{SigmaMellin}
\Sigma(V) \>=\> \int \frac{d\nu}{2\pi i \, \nu} \>\> e^{\nu V}\> 
  e^{-\cR[u]} \>,\qquad 
  e^{-\cR[u]} \>=\> \sum_{n}\int dw_n\> \prod_{i=1}^m \>u(k_i)\>.  
\end{equation}
For the $D$ distribution in \eqref{SD} we have to introduce two Mellin
transforms:
\begin{equation}
  \label{SDM}
\Sigma(A,\bar A) \>=\> 
\int \frac{d\nu_1}{2\pi i \, \nu_1} \>\> e^{\nu_1 A}\> 
\int \frac{d\nu_2}{2\pi i \, \nu_2} \>\> e^{\nu_2 \bar{A}}\>
e^{-\cR[u]} \>.
\end{equation}
The radiator $\cR$ is a functional of the source $u(k$) and
has the following two-loop expression: 
\begin{multline}\label{R2loop}
\cR[u]=
{4C_F} \int_{k_t^2}^1 \frac{d\al}{\al} \frac{d^2k_{t}}{\pi k_{t}^2}
\>\left(\frac{\as(0)}{4\pi} +\chi(k_{t}^2)\right)\> [1-u(k)] \\
+4C_F \int
d\Gamma_2(k_1,k_2)\left(\frac{\as}{4\pi}\right)^2\frac1{2!}M^2(k_1,k_2)
\;[1-u(k_1)u(k_2)] \,.
\end{multline}
For the specific observables under consideration the source functions
are:
\begin{equation}
\begin{split}
\label{sources}
T:\qquad 
u(k)&= e^{-\nu\al}\Theta(\beta-\al)+
e^{-\nu\beta}\Theta(\al-\beta)\>, \\
D: \qquad  u(k)&= \Theta(\beta-\al)e^{-\nu_1\al}
            +\Theta(\al-\beta)e^{-\nu_2\beta} \>, \\
C:   \qquad   u(k)&= e^{-\nu c(k)}\,, \quad
              c(k)= 6\frac{\al\, \be}{\be+\al}\>, \\
B:  \qquad  u(k)&= e^{-\nu\sqrt{\al\beta}}
e^{i b\sqrt{\al\beta}\cos\Phi}\>.
\end{split}
\end{equation}
In the source for the distribution in total broadening, $B=B_T$, 
we have introduced (see~\cite{DLMSbroad}), 
in addition to the Mellin variable $\nu$ conjugate to the modulus of
the transverse momentum, 
the two-dimensional impact parameter $\vec{b}$ conjugate to
the vector transverse momentum, and we write 
$\vec{b}\vec{k}_t=b\sqrt{\al\be}\cos\Phi$.

In \eqref{sources} we have expressed the sources as a function of the
Sudakov components $\alpha$ and $\beta$,
$$
   u(k) \>\equiv \> u(\alpha,\beta)\>,
$$
where, for a massless parton, $\alpha\beta=k_t^2$.

\subsection{The radiator}

The various terms in $\cR$ are ill-defined and only the sum has
physical meaning. In what follows we cast $\cR$ as sum of three 
contributions which, for our observables, are collinear 
and infrared finite.
To this end we introduce a source corresponding to the parent 
gluon momentum $u(k_1+k_2)$ and split the source in the two-parton 
contribution on the second line of \eqref{R2loop} as
\begin{equation}
  \label{naivesplit}
   1-u(k_1)u(k_2) \>=\> \left[\,1-u(k_1+k_2)\,\right]
\>\>+\>\> \left[\, u(k_1+k_2)-u(k_1)u(k_2)\,\right].  
\end{equation}
The expression in the first square brackets defines the ``naive
contribution'' which treats the gluon decay inclusively. 
The second contribution to the radiator 
we refer to as the non-inclusive correction. 
The parent gluon momentum $k_1+k_2$ is ``massive'':
$$
u(k_1+k_2)\>\equiv\> u(\al,\be)\>, \qquad \al=\al_1+\al_2\>,\>\>
\be=\be_1+\be_2\>; \quad \al\be=k_t^2+m^2\>.
$$
We note that the use of the sum of 4-vectors of the offspring
parton momenta in \eqref{naivesplit} is somewhat arbitrary.  We could
have used any combination of the momentum components which satisfies
$v=v_1+v_2$ in the limits of collinear and/or soft branching.  The
non-inclusive correction would then change correspondingly.  In other
words, both the naive coefficient for the power correction and the
Milan factor depend on the definition of the ``naive approximation,''
while the final answer remains unambiguous.  Our prescription has the
advantage of preserving those very simple numerical coefficients that
are known from previous one-loop calculations.

Noticing that the variables $\vec{k}_{t1}$, $\vec{k}_{t2}$, $z$ and
$m^2$ are invariant with respect to Lorentz boosts in the longitudinal
direction, while $\al$ is not, we conclude that the invariant matrix
element $M^2$ does not depend on $\al$.  This allows us to perform the
$\al$-integration explicitly and represent the non-inclusive
correction in the factorised form:
\begin{equation}\label{Rnigen}
\cR_{ni}[u]\>=\>
4C_F\int dm^2
dk_t^2 \>\frac{d\phi}{2\pi}\> dz\>
\left(\frac{\as}{4\pi}\right)^2 
\frac1{2!} M^2\>\Om_{ni}\,,
\end{equation}
with the non-inclusive ``trigger function''
\begin{equation}
\label{nitrigger}
\Om_{ni}\>\equiv\> \int_{k_t^2+m^2}^1\frac{d\al}{\al}
\;[\, u(k_1+k_2)-u(k_1)u(k_2)\,]\,.
\end{equation}
The $m^2$-integral converges, in spite of the singular behaviour of
the matrix element, $M^2\propto 1/m^2$,   
because the trigger function $\Om_{ni}$ vanishes
in the collinear limit, $\Om_{ni}\propto \sqrt{m^2}$.  

The radiator then takes the form
  \begin{multline}
    \label{Rsplit}
  \cR[u]= \cR_{ni}\>+\>
4C_F\int_0^1\frac{dk_t^2}{k_t^2}\left(\frac{\as(0)}{4\pi}
  +\chi(k_t^2)\right)\Om_0(k_t^2) \\
  + 4C_F\int_0^1 dm^2\, dk_t^2\, dz\, \frac{d\phi}{2\pi} 
  \>\frac{M^2(k_1,k_2)}{2!} \Om_0(k_t^2+m^2)\>,    
  \end{multline}
where we have introduced the ``naive'' trigger function
\begin{align}
  \label{0trigger}
  \Om_0(k^2)\>&\equiv\> \int_{k^2}^1 \frac{d\al}{\al} 
  \left[\,1-u\left(\al,\,\be={k^2}/{\al}\right)\right].
\end{align}
The inclusive integral of the two-parton matrix element 
on the second line of \eqref{Rsplit} is given, at fixed $k_t$, 
by~\cite{DLMSthrust}
\begin{multline}
\label{M2int}
\int dm^2\> dz\>\frac{d\phi}{2\pi} 
\left(\frac{\as}{4\pi}\right)^2 \frac 1{2!} M^2(k_1,k_2) \\
=\int_0^\infty \frac{dm^2}{m^2(m^2+k_t^2)}
\left(\frac{\as}{4\pi}\right)^2
\left\{-\beta_0+2C_A\ln\frac{k_t^2(k_t^2+m^2)}{m^4}\right\},
\end{multline}
$$
 \be_0\>=\> \frac{11}{3}C_A -\frac23n_f\>,
$$
where, for $k_t^2\ll1$, 
we have replaced the actual upper limit of the $m^2$ integral, 
$m^2<1$, by $m^2<\infty$ since the integral is convergent in the
ultra-violet region.

The $m^2$-integral here is ill-defined since it diverges in the
collinear two-parton limit $m^2\to0$.  The quantity $\as(0)$ on the
first line of \eqref{Rsplit} is also ill-defined. It combines,
however, with the $\beta_0$--term in \eqref{M2int} to produce the
running coupling at the scale $k_t$ in the one-gluon emission, $\cR_0$
(see below).

The virtual correction $\chi(k_t^2)$ also contains a collinear
divergence.  It can be written down in terms of the dispersive
integral
\begin{equation}\label{chidef}
\chi(k_t^2)= \int_0^\infty \frac{d\mu^2 \>k_t^2}{\mu^2(k_t^2+\mu^2)}
\left(\frac{\as}{4\pi}\right)^2 \left\{
-2C_A\ln\frac{k_t^2(k_t^2+\mu^2)}{\mu^4}
\right\} + 2 \left( \frac{\as}{2\pi}\right)^2 \cS
\,,
\end{equation}
with $\cS$ a scheme dependent number.
In the physical scheme, in which the coupling is
defined as the intensity of soft gluon radiation~\citd{CMW}{DKT}, 
one has $\cS=0$.  In the $\MSbar$ scheme, for example we have
\begin{equation}
  \label{eq:Kdef}
  \cS\>=\> K\equiv
  \ca\left(\frac{67}{18}-\frac{\pi^2}{6}\right)-\frac{5}{9}n_f  \>. 
\end{equation}
The collinear divergences in $\chi$ 
at $\mu^2\to0$ and in the logarithmic term of \eqref{M2int} at
$m^2\to0$, cancel for the collinear/infrared-safe observables we are
dealing with.  The remaining correction, the ``inclusive correction''
$\cR_{in}$, is finite.

Thus the final result for the perturbative radiator 
$\cR$ at the two-loop level 
can be cast as a sum of three terms 
$$
 \cR \>=\> \cR_0 + \cR_{in} + \cR_{ni}\>.
$$ 
This regularisation procedure was introduced and discussed in detail 
for the case of thrust in~\cite{DLMSthrust}. 

In the following we study the structure of these contributions in
terms of the so-called effective coupling $\ae$
which was introduced in \cite{DMW}.
It is related with the standard QCD running coupling by the dispersive
integral 
\begin{equation}\label{asaedisp}
\frac{\as(k^2)}{k^2}\;=\;
\int_0^\infty{dm^2}\frac{\ae(m^2)}{(m^2+k^2)^2}\,.
\end{equation}
Its logarithmic derivative is the spectral density for $\as$;
perturbatively,
\begin{equation}
  \label{rg}
\frac{d}{d\ln m^2} \frac{\ae(m^2)}{4\pi}
\>=\>  -\be_0\left(\frac{\as}{4\pi}\right)^2 +\ldots \>, \qquad 
\ae(0)=\as(0)\>.
\end{equation}

\paragraph{Naive contribution.}
The first contribution is the so-called ``naive'' con\-tri\-bu\-tion
which emer\-ges from the inclusive treatment, that is when the
contributions to $V$ of gluon decay products are replaced by that of
the parent ``massive'' gluon.  It incorporates only the term
proportional to $\be_0$ in \eqref{M2int} which, together with $\as(0)$
in the Born term of the radiator, builds up the running coupling in
the effective one-gluon emission.  We have
\begin{equation}
\label{cR0}
\cR_0 \> \equiv\>
4C_F \int \frac{dm^2 dk^2_t}{k^2_t+m^2}
\left\{\frac{\as(0)}{4\pi}\delta(m^2)
-\frac{\beta_0}{m^2}\left(\frac{\as}{4\pi}\right)^2\right\}
\Om_0(k_t^2+m^2)\>.
\end{equation}
Using \eqref{rg} and integrating by parts we arrive at
\begin{equation}
\begin{split} 
\label{cRnaive}
\cR_0[u] \>&=\> \frac{C_F}{\pi}\int_0^1 dm^2 \ae(m^2)\>\frac{-d}{dm^2}
\int_0^1\frac{dk^2_t}{k^2_t+m^2} \>\Om_0(k_t^2+m^2) \\
 &=\> \frac{C_F}{\pi}\int_0^1 \frac{dm^2}{m^2} \ae(m^2)
\>\Om_0(m^2)\>.
\end{split}
\end{equation}

\paragraph{Inclusive correction.}
The second contribution is the ``inclusive correction'' which takes
into account the logarithmic piece in \eqref{M2int} dropped in the
naive treatment.  If massless and massive gluons contributed equally
to $V$, this logarithmic term would cancel completely against the
second loop virtual correction $\chi$ to one-gluon emission. The
mismatch is proportional to the difference of the massive and massless
trigger functions.  Using \eqref{rg} and integrating by parts we
obtain
\begin{equation}\label{cRin}
\begin{split}
\cR_{in}[u]
&=\frac{8C_FC_A}{\beta_0} \int \frac{dm^2}{m^2}
\frac{\ae(m^2)}{4\pi} \frac{d}{d\ln m^2}
\int \frac{dk^2_t}{k^2_t+m^2}
\;\ln\frac{k_t^2(k_t^2+m^2)}{m^4}\;
\Om_{in}(k_t^2,m^2)\,, \\
\Om_{in} &\equiv\>  \Omega_0(k_t^2+m^2)-\Omega_0(k_t^2)\>.
\end{split}
\end{equation}

\paragraph{Non-inclusive correction.}
Finally, the ``non-inclusive correction'' in \eqref{Rnigen} 
accounts for non-soft, non-collinear gluon decays into two partons 
$k_1$, $k_2$, which kinematical configurations are mistreated by the
``naive'' inclusive approach, as was pointed out 
by Nason and Seymour~\cite{NS}.     
Using \eqref{rg} and integrating by parts we obtain 
\begin{equation}
\label{cRni}
\cR_{ni}[u]\>=\>
\frac{4C_F}{\be_0} \int_0^1 \frac{dm^2}{m^2} \>\frac{\ae(m^2)}{4\pi}
\frac{d}{d\ln m^2} \left\{ \int_0^{2\pi}\frac{d\phi}{2\pi}\int_0^1 dz 
\int_0^1 dk_t^2 \> m^2\frac1{2!} M^2\>\Om_{ni} \right\},
\end{equation}
which form is suited for extracting the power correction.

In the following we discuss the three trigger functions, $\Om_0$,
$\Om_{in}$ and $\Om_{ni}$, and then the corresponding power terms,
$\delta\cR_0$, $\delta\cR_{in}$, $\delta\cR_{ni}$.  We first consider
the ``Linear'' observables $T,D$ and $C$, and then the special case of
broadening, $B$, (the ``Log-Linear'' observable).

\subsubsection{The trigger functions for the Linear observables 
($V=1\!-\!T,D,C$)}

\paragraph{Naive contribution.}
The ``trigger functions'' $\Om_0(m^2)$ for our observables are: 
\begin{equation}\label{TDCtriggs}
\begin{split}
T:\qquad 
\Om_0&= 2\int_0^{\ln m^{-1}} d\eta 
   \left(1-\exp\left\{-m\nu e^{-\eta}\right\}\right), \\
D: \qquad 
\Om_0&=
  \int_0^{\ln m^{-1}} d\eta 
   \left(1-\exp\left\{-m\nu_1 e^{-\eta}\right\}\right)
+  \left(1-\exp\left\{-m\nu_2 e^{-\eta}\right\}\right), \\
C:   \qquad   
\Om_0&= \int_{\ln m}^{\ln m^{-1}} d\eta \left(1-\exp\left\{
-m\nu\,\frac{3}{\cosh\eta}\right\} \right) .
\end{split}
\end{equation}
Each $\Om_0$ vanishes $\propto\sqrt{m^2}$ 
in the small-mass limit, thus ensuring the
convergence of the $m^2$--integration in (\ref{cRnaive},\ref{cRin}).

The main, perturbative, contributions to \eqref{cRnaive}--\eqref{cRni}
come from the logarithmic integration region $1/\nu^2< m^2< 1$. The
expression \eqref{cRnaive} for $\cR_0$ reproduces then the usual
double-logarithmic perturbative radiator, proportional to
$\as\ln^2\nu$. The inclusive and non-inclusive contributions
\eqref{cRin} and \eqref{cRni} produce two-loop subleading corrections
of the order of $\as^2\ln\nu$. Actually, the standard resummation
programme can only guarantee exponentiation of the terms that contain
$\as^n\ln^m\nu$, with $m\ge n$. Therefore, the perturbative
corrections to $\cR_{in}$ and $\cR_{ni}$ do not belong to the
``radiator'' and are taken care of by the exact two-loop calculation
(the perturbative matching procedure)~\cite{CTTW}.

The genuine non-perturbative contributions originate from the mass
integration region $m^2\ll 1/\nu^2$. 
In this limit, we can expand the exponents in the trigger functions 
\eqref{TDCtriggs} to first order in $m$ to obtain
$$
 \Om_0(m^2;V)= \delta\Om_0 + \cO{m^2}\>; \quad 
 \delta\Om_0^{(V)} \>=\>  \rho^{(V)}\cdot m\>,
$$
where the $\rho$-parameters are given by a standard rapidity integral
depending on the observable under consideration:  
\begin{equation}\label{rhofacdef}
\rho^{(T)}=2\,\nu\int_0^\infty d\eta\> e^{-\eta}=2\nu\>,
\quad \rho^{(D)}=\nu_1+\nu_2\>,\quad   
 \rho^{(C)}=\nu\,\int_{-\infty}^\infty d\eta \frac{3}{\cosh\eta}=3\pi\nu\>.
\end{equation}

\paragraph{Inclusive correction.}
In the small-$m$ limit we have ($V=1\!-\!T,D,C$)
\begin{equation}
  \label{dOmin}
  \delta\Om_{in}^{(V)}\>=\> \rho^{(V)}\cdot
  \left(\sqrt{k_t^2+m^2}-k_t\right) 
\end{equation}
with the $\rho$-factors as in \eqref{rhofacdef}.

\paragraph{Non-inclusive correction.} 
In the small-$m$ limit the trigger function $\Om_{ni}$ in
\eqref{nitrigger} gets simplified:
\begin{equation}
\label{Omnisimp}
\de\Om_{ni}^{(V)}=\rho^{(V)}\cdot(k_{t1}+k_{t2}-\sqrt{k_t^2+m^2})\,.
\end{equation}
In the linear approximation, each source $u(k)$ produces a {\em
transverse mass}\/ factor ($\sqrt{k_t^2+m^2}$ for a massive gluon,
and $k_{ti}$ for a massless parton), the rest being the universal
observable-dependent factor $\rho^{(V)}$ defined in \eqref{rhofacdef}.
The broadening case is analogous, as discussed below.

\subsubsection{The trigger functions for the Log-Linear observable
  ($V=2B$)} 
The broadening measure should be treated separately since the
large-distance contribution to $B$ is enhanced by a $\ln m^2$ factor
as compared with the ``Linear'' $T,D$ and $C$.

\paragraph{Naive contribution.}
In the case of broadening the source is independent of $\eta$ 
for fixed (and small) transverse momentum, or rather transverse
mass. 
As a result, an additional $\log$-factor originates from integrating over
the rapidity of a parton: 
\begin{equation}
\label{broadeta}
B:  \qquad  
\Om_0= \left(1-e^{-m\nu}J_0(mb)\right)\> 
\int_{\ln m}^{\ln m^{-1}} d\eta\>.   
\end{equation}
In the small-$m$ limit we obtain
$$
\delta\Om_0^{(B)} \>=\> \nu\, m\cdot \ln\frac1{m^2} \>.
$$
This result, however, can be improved by taking into account the
non-soft contribution to the gluon radiation probability. This can be
done by simply using the exact quark $\to$ quark $+$ gluon splitting
function instead of the soft $d\al/\al$ spectrum in \eqref{dwn}:
\begin{multline}
  \int_{\ln m}^{\ln m^{-1}} d\eta = 2\int_{\sqrt{k_t^2+m^2}}^1
  \frac{d\al}\al 
  \Longrightarrow \\
 2\int_{\sqrt{k_t^2+m^2}}^1 d\al\>\frac{1+(1-\al)^2}{2\>\al} \>=\> 
 \ln \frac1{k_t^2+m^2}- \frac32 +\cO{k_t^2+m^2}\>.
\end{multline}
With account of this correction, the small-$m^2$ limit of the
broadening trigger function becomes 
\begin{equation}
  \label{rhobroad}
 \delta\Om_0^{(B)} \>=\> \nu\, m\cdot
 \left(\ln\frac1{m^2}-\frac32\right)   
 \>=\> \nu\,m\cdot  \ln\frac{e^{-3/2}}{m^2}\>.
\end{equation}
Note that the $b$-space resummation does not affect the leading
power correction: the expansion of the exponent 
$\exp(i\vec{b}\vec{k}_t)$ in \eqref{sources} 
lacks a term linear in $k_t$ after integration over the azimuthal 
angle $\Phi$.
Thus, a small transverse momentum of a gluon, and not that of the
primary quark, is relevant for the $1/Q$ power contribution to 
the broadening.

\paragraph{Inclusive correction.}
Given the definition of the inclusive trigger function~\eqref{dOmin},
we substitute the naive trigger function for the broadening,
\eqref{rhobroad}, to obtain
\begin{equation}
  \label{dOmbroad}
  \delta\Om_{in}^{(B)}\>=\> \nu
\left[\, \sqrt{k_t^2+m^2}\ln\frac{e^{-3/2}}{k_t^2+m^2}
-k_t\ln\frac{e^{-3/2}}{k^2_t}\right].  
\end{equation}

\paragraph{Non-inclusive correction.}
From the definition of the non-inclusive trigger function,
\eqref{nitrigger}, for the broadening we derive
\begin{equation}\label{OmniB}
\de\Om_{ni}^{(B)}=\nu \left(k_{t1}+k_{t2}-\sqrt{k_t^2+m^2}\right)
\ln\frac{e^{-3/2}}{k_t^2+m^2}
\>.
\end{equation}

\section{Power corrections to the radiators}
To extract the leading power contributions we replace the effective
coupling $\ae(m^2)$ in the integrals for $\cR_0$ \eqref{cRnaive},
$\cR_{in}$ \eqref{cRin} and $\cR_{ni}$ \eqref{cRni} by its
non-perturbative component (``effective coupling modification'')
$\de\ae(m^2)$ and employ the small-$m^2$ limit of the corresponding
trigger functions, $\de\Om$.  

The answers will contain the non-analytic in $m^2$ moments
\cite{BBZ94} of the effective coupling modification,
\begin{equation} \label{aefmom}
  A_{2p,q} = \frac{C_F}{2\pi}
  \int_0^\infty \frac{dm^2}{m^2}\> (m^2)^p \ln^q \frac{m^2}{\mu^2} 
  \> \de\ae(m^2)\>, 
\end{equation}
with $\mu$ an arbitrary momentum scale. We remind the reader that the
moments analytic in $m^2$ vanish, that is for $p$ integer and
$q=0$~\cite{DMW,BB95}.

\subsection{Linear observables ($V=1\!-\!T,D,C$)}

\paragraph{Naive contribution.}
We start from $\de\cR_0$, the power correction to the naive
contribution:
\begin{equation}\label{R0pow}
\de\cR_0^{(V)}\>=\>
\frac{C_F}{\pi}\int \frac{dm^2}{m^2} \de\ae(m^2)\>
\de\Om_0^{(V)}\>=\> \rho^{(V)}\cdot \frac{2A_1}{Q}\>,
\end{equation}
where we have restored the dimensional factor $1/Q$.
Here $A_1$ is the $p=\half, q=0$ moment of the coupling
modification \eqref{aefmom}:
\begin{equation}\label{A1def}
  A_1 \equiv A_{1,0} = \frac{C_F}{2\pi}
  \int_0^\infty \frac{dm^2}{m^2}\> m \> \de\ae(m^2)\>.
\end{equation}

\paragraph{Inclusive contribution.}
We have to consider the following integral in \eqref{cRin}:
\begin{multline}\label{jk}
\frac{d}{d\ln m^2} \left\{ \int_0^\infty \frac{dk^2_t}{k^2_t+m^2}
\;\ln\frac{k_t^2(k_t^2+m^2)}{m^4}\;
\de \Om_{in}^{(V)} \right\} \\
=
\rho^{(V)}\cdot
\frac{d}{d\ln m^2} \left\{ \int_0^\infty \frac{dk^2_t}{k^2_t+m^2}
\;\ln\frac{k_t^2(k_t^2+m^2)}{m^4}\>
\left(\sqrt{k_t^2+m^2}-k_t\right)\right\}. 
\end{multline}
We have replaced here the actual upper limit of the $k_t^2$
integration, $Q^2=1$, by $\infty$ since the correction term is
proportional to $m^2\ln m^2$ 
and does not produce the leading $1/Q$ power contribution.
Introducing the dimensionless integration variable $x=k_t/m$ 
we obtain the ``inclusive'' power correction in the form
\begin{equation}
\de\cR_{in}^{(T,D,C)}=\de\cR_0^{(T,D,C)}\>r_{in}\,,
\end{equation}
with 
\begin{equation}
  \label{rin}
  r_{in} \>=\>  \frac{2C_A}{\be_0} \int_0^\infty
\frac{x\>dx}{(1+x^2)(\sqrt{1+x^2}+x)}\;\ln(x^2(1+x^2))
\>=\> 3.299 \, \frac{C_A}{\be_0}\>.
\end{equation}

\paragraph{Non-inclusive contribution.}
To calculate the $1/Q$ power term in the {\it non-inclusive}\/
contribution in \eqref{cRni} we have to consider the multiple 
integral of the two-parton decay probability
\begin{equation}
\rho^{(V)}\frac{d}{d\ln m^2} 
\left\{ \int_0^{2\pi}\frac{d\phi}{2\pi}\int_0^1 dz 
\int_0^1 dk_t^2 \> m^2\frac1{2!} M^2\> 
\left(k_{t1}+k_{t2}-\sqrt{k_t^2+m^2}\right)\right\}.
\end{equation}
The linear mass-dependence can be factored out, leaving the convergent
dimensionless integral which was calculated in~\cite{DLMSthrust}. 
The result reads
\begin{equation}\label{niTDC}
\de\cR_{ni}^{(T,D,C)}
\>=\>\de\cR_0^{(C,D,C)}\;r_{ni}\>, \qquad 
r_{ni}\>=\> {2\be_0^{-1}}\left( -1.227C_A\>+\> 0.365C_A\>-\>
  0.052n_f\right),  
\end{equation}
where the three terms originate respectively from the soft-gluon,
hard-gluon and quark parts of the matrix element
(see~\cite{DLMSthrust}).

\paragraph{The final results for $T,D,C$.}
Finally, the leading power contributions to the radiators of the $T,D$
and $C$ distributions are:
\begin{equation}
\label{finallinear}
\de\cR^{(T,D,C)}\>=\> \rho^{(V)}\frac{2A_1}{Q}\,\cM
\>=\> \de\cR_0^{(T,D,C)}\>\cM,
\end{equation}
with the $\rho$ factors given in \eqref{rhofacdef}.  This answer
differs from the naive predictions $\cR_0^{(V)}$ defined in
\eqref{R0pow} by the {\em universal}\/ ``Milan factor'',
\begin{equation}\label{cMdef}
\begin{split}
\cM\>&=\>1+ r_{in}+r_{ni} \>=\>
1+{\be_0^{-1}}\left( 1.575C_A\>-\> 0.104n_f\right) \\ 
\>&=\> 1.490\>(1.430) \quad \mbox{for}\>\> n_f=3\>(0)\>.
\end{split}
\end{equation}
We combine the Milan factor with the $A_1$ moment to define a small
parameter 
\begin{equation}
  \label{cPdef}
  \cP\>\equiv\> \frac{2A_1}{Q}\cM\>. 
\end{equation}
The non-perturbative contribution to the radiator becomes 
$\de\cR^{(V)}=\rho^{(V)}\cP$.

\subsection{The case of Broadening}
The answer for broadening\footnote{In this revised version we wish to
  point out that the analysis of the jet-broadenings presented here is
  superseded by that given in Eur. Phys. J. direct C5 (1999) 1. While
  the previous section \emph{has} been updated to account for a
  mistake of a factor of two for the non-inclusive pieces in the
  original version of the paper, this section has not been updated
  (this would simply involve the inclusion of an extra factor of two
  in $r_{ni}'$) since it would in any case not represent the full
  answer.}  contains two moments: the $\ln Q$--enhanced term
proportional to $A_1$ and a new log-moment \eqref{aefmom} with
$p=\half, q=1$:
\begin{equation}
  \label{A11}
 A_{1,1}\equiv A_1' \>=\>  
  \frac{C_F}{2\pi}
  \int_0^\infty \frac{dm^2}{m^2}\> m
  \ln \frac{m^2}{\mu^2} \> \de\ae(m^2)\>.
\end{equation}

\paragraph{Naive contribution.}
The naive power contribution to the $B$--radiator reads, see
\eqref{cRnaive}, \eqref{rhobroad},
\begin{equation}
  \label{cRnaiveB}
  \de\cR_0^{(B)} \>=\> \nu \cdot
  \left[\,\frac{2A_1}{Q}\left(\ln\frac{Q^2}{\mu^2}-\frac32\right)
   - \frac{2A_1'}{Q}\,\right] .
\end{equation}
Since the 
value  $A_{1}'$ is not apriori known, we can
eliminate it by choosing $\mu$ which then
becomes a new non-perturbative parameter.

\paragraph{Inclusive contribution.}
For the $B$-distribution \eqref{jk} becomes
\begin{multline}
\nu \,
\frac{d}{d\ln m^2} \left\{ \int_0^\infty \frac{dk^2_t}{k^2_t\!+\!m^2}
\;\ln\frac{k_t^2(k_t^2\!+\!m^2)}{m^4}\> 
\left(\sqrt{k_t^2\!+\!m^2}\ln\frac{Q^2{e^{-3/2}}}{k_t^2\!+\!m^2}
-k_t\ln\frac{Q^2{e^{-3/2}}}{k_t^2}\right)\right\}.
\end{multline}
The answer reads, see \eqref{dOmbroad}
\begin{equation}
\de\cR_{in}^{(B)} \>=\>\nu\left\{ r_{in}\frac{2A_1(\ln Q^2/\mu^2-3/2-2)
    -2A_1'}{Q} + r'_{in}\frac{2A_1}{Q}\right\},
\end{equation}
where
\begin{equation}
  \label{rinprime}
r'_{in} = \frac{2C_A}{\be_0}
\int_0^\infty \frac{x\>dx}{(1+x^2)}\left[\,
x\>\ln x^2-\sqrt{1\!+\!x^2}\>\ln(1\!+\!x^2)\,\right]\ln(x^2(1\!+\!x^2))
= -22.751 \frac{C_A}{\be_0}\,.
\end{equation}

\paragraph{Non-inclusive contribution.}
The non-inclusive power correction to the broadening radiator is
obtained by using \eqref{OmniB}, the small-$m$ part of the trigger
function, in \eqref{Rnigen}:
\begin{equation}
\de\cR_{ni}^{(B)} \>=\>\nu\left\{ r_{ni}\frac{2A_1(\ln Q^2/\mu^2-3/2-2)
    -2A_1'}{Q} + r'_{ni}\frac{2A_1}{Q}\right\},
\end{equation}
where $r_{ni}'$ is given by the dimensionless integral similar to that
for $r_{ni}$ in \eqref{niTDC} but with an additional logarithmic
factor:
\begin{equation}
  \label{niprimeB}
  r'_{ni}\>=\> (4\be_0)^{-1}
  \int\frac{d\phi}{2\pi} \int dz\int \frac{d^2k_t}{\pi}\> (m^2 M^2) 
 \frac{k_{t1}+k_{t2}-\sqrt{k_t^2+m^2}}{m}\>  \ln \frac{m^2}{k_t^2+m^2}.
\end{equation}
Its numerical value is
\begin{equation}
  \label{rniprime}
  r_{ni}'\>=\> {\be_0^{-1}}
  \left( 4.808C_A\>-\> 0.884C_A\>+\>
    0.116n_f\right)\,,
\end{equation}
where the three terms, as before, correspond to the soft-gluon,
hard-gluon and quark contributions.

\paragraph{The final result for $B$.}
The full power term in the broadening radiator is obtained by
combining the three contributions. One reconstructs the Milan factor
and finds
\begin{equation}
  \label{broadpower}
\de\cR^{(B)}=  
2\nu\cdot\frac{2\cM}{Q}\left[\,A_1\left(\ln\frac{Q}{\mu} -\xi\right)
-\left(\half{A_1'}+{A_1}\right)  \,\right],
\end{equation}
where
\begin{equation}
  \label{xidef}
 \xi \>=\>  \frac34 - \frac{2+r'_{in}+r'_{ni}}{2\cM}
= 1.692+ \frac{0.0765\, n_f}{C_A/3-0.0392n_f} 
= 1.930\>
(1.692)\qquad \mbox{for}\> n_f=3\>(0)\>.
\end{equation}
Here we have introduced the combination of
the moments, $\half A_1'+A_1$, which will be of convenience later on,
see section~\ref{sec:merge}.

Defining also the parameter $\cP'$, we can write
\begin{equation}
  \label{cPprdef}
  \de\cR^{(B)}=  
2\nu\cdot\left[\,\cP\left(\ln\frac{Q}{\mu} -\xi\right)
-\cP' \,\right], \qquad 
  \cP' \>\equiv\> \frac{A_1'+2A_1}{Q}\cM\>.
\end{equation}
The new unknown parameter, $\cP'$ can be traded for the scale of
the logarithm:
\begin{equation}
  \label{finalloglinear}
\de\cR^{(B)} 
\>=\> 2\nu\cdot \cP\,\ln\frac{Q}{Q_{B}}\>.
\end{equation}
The scale $Q_B$ is observable-dependent and reads
\begin{equation}\label{QBPprime}
 \ln\frac{Q_{B}}{\mu} \> =\>  \frac{\half A_1'+A_1}{A_1} + \xi
\>=\> \frac{\cP'}{\cP}+\xi\>.   
\end{equation}

\section{Power effects in the distributions}

The leading power contributions to the radiators are proportional to
the corresponding Mellin variable $\nu$, see~\eqref{finallinear},
\eqref{finalloglinear}.  Inserting this into \eqref{SigmaMellin} one
obtains a {\em shift}\/ of the corresponding perturbative
distribution~\cite{DW97} proportional to the $\cP$ parameter
\eqref{cPdef}.

We arrive, for small $V=1\!-\!T,D,C,B$ at
\begin{eqnarray}
\label{final}
\frac{d\sigma}{ dT}(1\!-\!T) &=&\left[\,\frac{d\sigma}{ dT}
\left(1-T-2\cP\right)\,\right]_{\PT}   \\
\label{AbarAdist}
\frac{d\sigma}{ dA\,d\bar{A}}(A,\bar{A})
&=& \left[\,\frac{d\sigma}{dA\,d\bar{A}}
\left(A-\cP,\,
\bar{A}-\cP\right)\,\right]_{\PT}  \\
\frac{d\sigma}{ dC}(C)
&=&  \left[\,\frac{d\sigma}{dC} 
\left(C-3\pi\cP\right)\,\right]_{\PT}  \\
\label{totalbroaddist}
\frac{d\sigma}{ dB}(B)
&=& \left[\,\frac{d\sigma}{ dB}
\left(B-\cP\ln({Q}/{Q_{B}})\right)\,\right]_{\PT}.
\end{eqnarray}
These expressions are valid as long as $V$, though numerically small,
stays at the same time much larger than $\lqcd/Q$, that is $V\gg \cP$.
The kinematical region $V<\cP$ is dominated by confinement
physics: the jets get squeezed down to small-mass (exclusive)
hadron systems. 

By expanding the general answer for the distributions in the Taylor
series in $\ln V$, 
\begin{equation}\label{fVexp}
  f(V)\>=\> f_{\PT}(V-c_V\cP) \>=\> f_{\PT}(V)
         -\frac{c_V\cP}{V} f'_{\PT}(V) +
         \half\left(\frac{c_V\cP}{V}\right)^2
          \left(f''_{\PT}-f'_{\PT}\right)  +\ldots\>, 
\end{equation}
with $g'(V)\equiv dg/d\ln V$, 
we observe that the true expansion parameter is $\cP/V$, rather than
$\cP$. 
Such an enhancement at small $V$ 
is a common feature of the power contributions. 
For example, the $1/Q^2$ power terms in DIS structure
functions and $e^+e^-$ fragmentation functions are relatively 
enhanced near the phase space boundary as $(1-x)^{-1}$, where $x$ is a
usual Bjorken (DIS) or Feynman ($e^+e^-$) variable~\cite{DMW}.

Strictly speaking, within our approach only the leading power
correction was kept under control.  Therefore we should have dropped
the rest of the series, starting from the quadratic term
$\cO{(\cP/V)^2}$.  In spite of this we prefer to present the result in
terms of the shifted perturbative spectra, because this form is easier
to implement practically.  The subleading power contributions become
comparable with the leading one for $V \ga \cP$ and remain to be
studied.

\subsection{Jet mass distribution}
The invariant jet mass distribution(s) can be derived from the double
differential distribution \eqref{AbarAdist} with use of the relation 
\eqref{SD}.
Generally speaking, the relation between the Sudakov variables $A$,
$\bar{A}$ and the jet masses, $M_L^2$, $M_R^2$, is non-linear:
\begin{align}
  dA\>d\bar{A} &= \frac{dM_R^2\, dM_L^2}{J}\>, \\
J &\equiv 1-A-\bar{A}= {\sqrt{1-2(M_R^2+M_L^2)+
(M_R^2-M_L^2)^2}}\>=\> 1+\cO{M^2} \>.
\end{align}
The relations inverse to \eqref{M2AB} are
\begin{align} 
 A &= M_R^2 \left(1+M_L^2\cdot\eps\right)\, \quad 
 \bar{A} = M_L^2 \left(1+M_R^2\cdot\eps\right)\; \\
 \eps &\equiv \frac{2}{1-M_R^2-M_L^2 +J}\>=\> 1+\cO{M^2}.
\end{align}
Calculating the mass distribution in terms of the shifted $A$,
$\bar{A}$ perturbative spectrum in \eqref{SD} amounts to shifting the
$M^2$ values according to
$$
 \frac{d^2\sigma}{dM_R^2\,dM_L^2}(M_R^2,M_L^2) \>=\> 
\left[\, \frac{d^2\sigma}{dM_R^2\,dM_L^2}
(M_R^2-\cP J+\cP^2,\,M_L^2-\cP J+\cP^2)\,\right]_{\PT}\>.
$$
The quadratic terms should be dropped within our accuracy. 
As for the linear power shift, in the small-mass limit,
$M_R^2,M_L^2\ll 1$, we can approximate $\cP J\simeq \cP$. 
Therefore, in the small $M^2$ region we can simplify the answer as
\begin{equation}\label{dMdis}
 \frac{d^2\sigma}{dM_R^2\,dM_L^2}(M_R^2,M_L^2) \>=\> 
\left[\, \frac{d^2\sigma}{dM_R^2\,dM_L^2}
(M_R^2-\cP,\,M_L^2-\cP)\,\right]_{\PT}\>.  
\end{equation}
Notice, however, that in the calculation of the {\em mean}\/ 
squared invariant mass(es) the Jacobian factor will induce a
perturbative correction to the non-perturbative shift of the order of 
$\cP\cdot \as(Q^2)$, since 
$\left\langle J\right\rangle 
\sim 1+\left\langle M^2\right\rangle = 1+\cO{\as}$.

\subsection{Power effects in whole-event and single-jet shapes}
Some shape observables characterise a single jet rather than the whole
(2-jet) ensemble. Among these are such characteristics as the
heavy-jet mass, the wide-jet broadening, etc., to be compared with the
mean jet mass, and the total $B$. 
Within the first order analysis aiming at power effects, that very
gluon that induced the power-behaving shift in the jet distribution,
was making the jet under consideration ``heavier'' (``wider'') than
its partner jet devoid of any radiation. This resulted in a strange
unphysical picture in which all the non-perturbative  
effects were contained in the heavy/wide jet alone~\cite{Web94}. A
proper treatment of the heavy-light relationship has been given by
Akhoury and Zakharov in \cite{AZ}.

To illustrate the solution of the puzzle, let us take the total and
heavy-jet mass distributions as an example.
For the distribution in $M_T^2=M_R^2+M_L^2$ we immediately obtain the 
{\em doubled shift}\/ of the corresponding perturbative expression, 
as compared with \eqref{dMdis}:
\begin{equation}
\begin{split}
 \frac{d\sigma}{d M_T^2}(M_T^2) &\equiv
 \int_0^{\infty} dM_R^2 \int_0^{\infty} dM_L^2 \>
 \delta(M_T^2-M_R^2-M_L^2)\>
 \left[\, \frac{d^2\sigma}{dM_R^2\,dM_L^2}
(M_R^2-\cP ,\,M_L^2-\cP)\,\right]_{\PT} \\
&= \int\int dM_1^2 \, dM_2^2 \>\delta(M_T^2-2\cP -M_1^2-M_2^2)\>
 \left[\, \frac{d^2\sigma}{dM_R^2\,dM_L^2}
(M_1^2,M_2^2)\,\right]_{\PT} \\
&\equiv \left[\, \frac{d\sigma}{d M_T^2}(M_T^2-2\cP)\,\right]_{\PT}.
\end{split}
\end{equation}
To calculate the {\em heavy-jet}\/ mass distribution one needs to
perform the integration over the smaller of the two jet masses (for
definiteness, we take $M_H=M_R>M_L$ and double the answer):
\begin{equation}
\begin{split}
 \frac{d\sigma}{d M_H^2}(M_H^2) &\equiv
2 \int^{M_H^2} dM_L^2 
 \left[\, \frac{d^2\sigma}{dM_R^2\,dM_L^2}
(M_H^2-\cP ,\,M_L^2-\cP)\,\right]_{\PT} \\
&=2 \int^{M_H^2-\cP} dM_1^2 
 \left[\, \frac{d^2\sigma}{dM_R^2\,dM_L^2}
(M_H^2-\cP ,\,M_1^2)\,\right]_{\PT} \\
&\equiv  \left[\, \frac{d\sigma}{d M_H^2}(M_H^2-\cP) \,\right]_{\PT} .
\end{split}  
\end{equation}
Recall that the perturbative answer for the integrated 
heavy-jet mass spectrum is essentially the {\em squared}\/ 
single-jet distribution~\cite{CTTW},
$$
\Sigma_H(M_H^2) = \left( \Sigma_{\mbox{\scriptsize
      one-jet}}(M_H^2)\right)^2,
$$
where  $\Sigma_{\mbox{\scriptsize one-jet}}$ describes a
single-inclusive distribution in, say, $M_R^2$, integrated over
$M_L^2$ without constraint.  This corresponds to putting
$\nu_2\!=\!0$ in \eqref{SDM} and \eqref{sources}. 
Thus we conclude that while the total-mass distribution acquires the
doubled shift, $2\cdot\cP$, 
($\nu_1\!=\!\nu_2$ in the Mellin transformation language), 
the heavy-mass spectrum is sensitive to confinement effects in a
single jet only ($\nu_2\!=\!0$), 
and therefore the corresponding perturbative
expression gets shifted by $1\cdot \cP$.

The same pattern holds for the total- versus wide-jet broadening:
\begin{equation}
\frac{d\sigma}{ dB_W}(B_W)
\>=\> \left[\,\frac{d\sigma}{ dB_W}
\left(B_W-\half\cdot\cP\ln({Q}/{Q_{B}})\right)\,\right]_{\PT},  
\end{equation}
to be compared with the twice larger shift in the total broadening
spectrum \eqref{totalbroaddist}.

\subsection{Energy-Energy Correlation}
At the perturbative level, the gluon-quark and quark-quark
contributions to the energy-energy correlation function $\EEC(\chi)$
defined in \eqref{EECdef} are comparable.  As far as the power effects
are concerned, the leading $1/Q$ contribution to EEC originates from
the correlation between one of the primary quarks and a very soft
gluon, $\EEC_{qg}^{\NP}$. The gluon has an energy of order of the
confinement scale, $\omega\sim k_t\sim \lqcd$.  It is easy to see that
this power correction to EEC acquires the same universal Milan factor
as the $T,D,C$ and $B$ distributions:
\begin{equation}
  \label{EECpow}
  \EEC(\chi) \>=\> \EEC^{\PT}(\chi) + \frac{4}{\sin^3\chi}\>
  \frac{A_1\,\cM}{Q}\>=\> \EEC^{\PT}(\chi) +\frac{2\cP}{\sin^3\chi} \>.
\end{equation}
While we will not present here explicit results for the quark-quark
contribution, $\EEC_{qq}^{\NP}$,
it is straightforward to show that it is of the form
$$
\EEC_{qq}^{\NP} \sim \frac{A_{2,1}}{Q^2\sin^4 \chi}\,.
$$
Both the quark-gluon and quark-quark contributions are more
singular at small $\sin\chi$ than the perturbative part of $\EEC$:
\begin{equation} 
  \label{EECNPtoPT}
 \frac{ \EEC_{qg}^{\NP}(\chi)}{ \EEC^{\PT}(\chi)} 
 \>\sim\> \frac{A_1}{Q\,\sin\chi} \>, \qquad
 \frac{ \EEC_{qq}^{\NP}(\chi)}{ \EEC^{\PT}(\chi)} 
 \>\sim\> \frac{A_{2,1}}{Q^2\,\sin^2\chi} \>.
\end{equation}
This, as we have discussed above, is a standard feature of 
power-suppressed contributions. The power-suppressed contribution from
the gluon-gluon correlation may contribute as much as the quark-quark
component, but it requires more study.

In the region of very small $\sin\chi$ values, all-order effects in
the perturbative quark-quark correlation become essential, which makes
the perturbative correlation finite in the back-to-back configuration,
that is for $\chi = \pi$ \citd{PP}{RW}.  Qualitatively, what happens
is that the quark and antiquark are not exactly back-to-back, with the
typical acollinearity angle, $\bar\chi$, being given by 
\begin{equation}
  \label{chibar}
  \sin \bar\chi = (Q \bar b)^{-1}\,.
\end{equation}
Here $\bar b = \bar b(Q)$ is the characteristic impact parameter which
determines the perturbative double-logarithmic distribution. A naive
estimate gives:
\begin{equation}
  {\bar{b}}(Q) \>\sim\> \frac1{Q} \cdot 
 \left(\frac{Q}{\lqcd}\right)^{\be_0/(\be_0+4C_F)}
  \>\propto\> Q^{-0.372}\qquad
\mbox{for}\>\> n_f=3\,.
\end{equation}
The acollinearity of the quarks leads to the freezing of the
singularities in \eqref{EECNPtoPT} when $\pi-\chi\la \bar\chi$.
For the back-to-back correlation we therefore expect
\begin{equation} 
  \label{EECpi}
 \frac{ \EEC_{qg}^{\NP}(\pi)}{ \EEC^{\PT}(\pi)} 
 \>\sim\> A_1 \> \bar b(Q) \>, \qquad
 \frac{ \EEC_{qq}^{\NP}(\pi)}{ \EEC^{\PT}(\pi)} 
 \>\sim\> A_{2,1} \>\bar{b}^{\,2}(Q)\,.
\end{equation}
We expect the same fractional powers of $Q$ for the non-perturbative
corrections to the value of the energy-weighted particle flow in the
current fragmentation region in DIS, in the photon direction
($q_t=0$).  In addition, for the value of $d\sigma/dM^2\,dq_t^2$, the
height of the plateau in the transverse momentum distribution of
Drell-Yan pairs ($W$, $Z$ bosons) at $q_t=0$, we expect a contribution
which is quadratic in $\bar b(Q)$,
$$
\left. \frac{d\sigma}{dQ^2 dq_t^2} \right|_{q_t=0}
=
\left[ \left. \frac{d\sigma}{dQ^2 dq_t^2} \right|_{q_t=0}\right]_{\PT} 
\exp\left\{
  \half A_{2,1}\cdot  \bar{b}^{\,2}(Q) \right\}.
$$
A more detailed analysis reveals that the estimates for the
values of the exponents are too naive, but that non-integer power
effects are present \cite{DMW2000}.

\section{Merging the PT and NP contributions}
\label{sec:merge}
We now discuss how to obtain the full expressions for the
distributions by merging the perturbative and the non-perturbative
contributions.  We follow the procedure suggested in~\cite{DW}.  The
answer we shall arrive at will differ, however, from that of~\cite{DW}
by a factor normalising the power terms.  Part of it is the Milan
factor we have discussed above.  As we shall shortly see, an
additional factor, $2/\pi$, arises due to the ``translation'' from the
effective coupling $\ae$ (which we use to trigger the large-distance
power contributions within the dispersive approach of \cite{DMW}) to
the standard QCD coupling $\as$.

For  the radiators of the $V=1\!-\!T,D,C$ 
distributions which contain $1/Q$ power terms
we have obtained
\begin{equation}
\label{cRPTNP}
  \cR(Q^2) = \cR^{\PT}(Q^2) + \rho^{(V)}\cP\>, 
\end{equation}
where $\cP$ is given in \eqref{cPdef} in terms of the moment 
$A_1$ of $\delta\ae$.

This representation is symbolic.  Indeed, its perturbative part is
given by an expansion which is factorially divergent.  Meanwhile, the
non-perturbative ``effective coupling modification'', $\delta\ae$,
also implicitly contains an ill-defined all-order subtraction of the
pure perturbative part off the full effective coupling $\ae$.

These two problems are of a similar nature. 
The uncertainty can be resolved at the price of introducing an
``infrared matching scale'', $\mu_I$. 
To this end we represent the answer as a sum of three terms:
\begin{equation}
\label{3tsplit}
\cR(Q) = \cR^{\PT}(Q;N) + \cR^{\NP}(Q,\mu_I) - \cR^{\merge}(Q,\mu_I;N)
\>+\> \cO{\as^{N+1}}\>,
\end{equation}
with $N$ the order at which the perturbative expansion is truncated. 

The factorially growing  PT-expansion terms (infrared
renormalons) in the first (PT) and the last (``merging'')
pieces in \eqref{3tsplit} cancel. 
The $\mu_I$-dependence in the second and the third
contributions approximately cancels as well, the cancellation
improving for larger $N$.

The structure of \eqref{3tsplit} is general; the same approach can be 
applied to incorporate the leading power contributions into different
observables $p,q$.  We proceed with
the radiators for $V=1\!-\!T,D,C$ distributions ($p=\half, q=0$) taken
as an example of the general matching procedure~\eqref{3tsplit}.  To
this end we split the running coupling, formally, into two pieces,
$$
\as(k^2) \>=\> \as^{\mbox{\scriptsize PT}}(k^2)\>+\>
\as^{\mbox{\scriptsize NP}}(k^2)\>.
$$
Here $\as^{\PT}(k^2)$ should be understood not as a mere function
but rather as an operational procedure: its momentum integrals should
be evaluated {\em perturbatively}, that is order by order in
perturbation theory as a series in $\as=\as(Q^2)$.  Its
non-perturbative counterpart, $\as^{\NP}(k^2)$, corresponds to the
$\de \ae(m^2)$ part of the effective coupling. It
is supposed to be a
rapidly falling function in the {\em ultra-violet}\/
region~\citd{DMW}{Grunberg}.  In the {\em infrared}\/ momentum region
neither of the two is a well-defined object.  However, the full physical
coupling $\as(k^2)$ is supposed to be finite down to $k^2\to0$. Its
infrared momentum integrals will play the r\^ole of phenomenological
parameters. To define these parameters we first use the formal
dispersive relation \eqref{asaedisp} for $\as^{\NP}$ and its partner,
$\delta\ae$, to deduce
\begin{equation}
  \int_0^\infty dk\>  \as^{\mbox{\scriptsize NP}}(k^2)
   \>=\> \frac{\pi}{4}\> \int_0^\infty \frac{dm^2}{m^2}\> m\>
   \de\ae(m^2)\>.
\end{equation}
Then, for $A_1$ we have
\begin{equation}
\label{A1cut}
  A_1 = \frac{2C_F}{\pi^2} \int_0^\infty dk\> \as^{\NP}(k^2)
\>=\> \frac{2C_F}{\pi^2} \int_0^{\mu_I} dk\> \as^{\NP}(k^2)
+ \cO{\frac{\mu_I}{Q}\as^{\NP}(\mu_I^2)}\>.
\end{equation}
Here we have introduced a scale $\mu_I$ above which the coupling is
well matched by its logarithmic perturbative expression, so that the
error proportional to $\mu_I\as^{\NP}(\mu_I^2)$ induced by truncation
can be neglected.  Now we substitute
$$ 
\as^{\NP}(k^2) \>=\> \as(k^2) - \as^{\PT}(k^2) 
$$ 
into \eqref{A1cut} and approximate $A_1$ as
\begin{equation}
\label{A1s}
  A_1  \simeq \frac{2\cf}{\pi^2}\int_0^{\mu_I} dk \>\as(k^2)
  -\frac{2\cf}{\pi^2}\int_0^{\mu_I} dk \>\as^{\PT}(k^2) \>.
\end{equation}
The first contribution can be expressed in terms of a
($\mu_I$-dependent) phenomenological parameter $\alpha_0$,
\begin{equation}
  \label{a0def}
   \int_0^{\mu_I} dk\> \as(k^2) \>\equiv\> \mu_I \cdot \alpha_0(\mu_I)\>.
\end{equation}
Thus, the power contribution in 
\eqref{cRPTNP} splits into two terms:
\begin{equation}
\label{NPsplitintwo}
  \de \cR^{(V)} = \rho^{(V)}\,\cP \>=\> \cR^{\NP}(Q,\mu_I) 
 - \cR^{\merge}(Q,\mu_I)\>.
\end{equation}
The first term is the integral over the infrared region of the full
coupling,
\begin{equation}
  \cR^{\NP}= \rho^{(V)}\>\frac{4C_F}{\pi^2}\>\cM\>
  \frac{\mu_I\,\alpha_0}{Q}\>.
\end{equation}
The subtraction  term is given by the integral of the perturbative
coupling,
\begin{equation}
  \cR^{\merge}(Q,\mu_I) \>=\>  \rho^{(V)}\frac{4C_F}{\pi^2}\>
\frac{\cM}{Q} \>\int_0^{\mu_I} dk\> \as^{\PT}(k^2)\>.
\end{equation}
This perturbative subtraction contribution 
is obtained by substituting for $\as^{\PT}(k^2)$ its 
perturbative series,
\begin{equation}
  \label{asPT}
  \as^{\PT}(k^2) = \sum_{\ell=1}^\infty \as^\ell\> P_{\ell}\>\left(\ln
  \frac{\mu_R}{k}\right)\>,  
  \qquad \as\equiv\as(\mu_R^2)\>.
\end{equation}
Here $\mu_R$ is the renormalisation scale which hereafter we set equal
to $Q$.  The representation \eqref{NPsplitintwo} is still symbolic,
since the all-order perturbative subtraction diverges.  Indeed,
integrating \eqref{asPT} term by term one finds 
\begin{equation}
    \int_0^{\mu_I} dk\> \as^{\PT}(k^2) \>=\> \mu_I\cdot \as
\sum_{\ell=0}^\infty \left(\be_0\frac{\as}{2\pi}\right)^\ell C_\ell\>,
\end{equation}
with $C_\ell$ being factorially growing coefficients. 
This behaviour is general. Taking, for the sake of illustration,  
the one-loop coupling, in which case \eqref{asPT}
is a simple geometric series, 
\begin{equation}
  \label{asPTone}
  \as^{\PT}(k^2) =  \as\sum_{\ell=0}^\infty \left(\beta_0\frac{\as}{2\pi}
\ln\frac{Q}{k}\right)^\ell\>,
 \quad \beta_0=\frac{11N_c}{3}-\frac{2n_f}{3}\>,
\end{equation}
one obtains
\begin{equation}
  \label{Cell}
 C_\ell = \frac{1}{Q}\int_0^{\mu_I}dk \left(\ln\frac{Q}{k}\right)^\ell
 = \int_{\ln (Q/\mu_I)}^\infty dt\> e^{-t}\> t^\ell \>, 
\qquad t=\ln \frac{Q}{k}\>.
\end{equation}
For finite $\ell$ we have $C_\ell \simeq \left(\ln
  ({Q}/{\mu_I})\right)^\ell$, while for $\ell\gg\ln(Q/\mu_I)$ the
coefficients start to grow as $C_\ell\sim\ell!$ (infrared renormalon).

The perturbative part of the answer, $\cR^{\PT}$ in \eqref{cRPTNP},
bears the same (infrared renormalon) divergence.  As we have
anticipated above, the factorial behaviours of the coefficients of
these two perturbative expansions cancel.  Indeed the corresponding
coefficients in $\cR^{\PT}$ for large $\ell$ are given by integrals
identical to \eqref{Cell} but with the upper limit taken as
$k\!\le\!Q$ ($t\!\ge\!0$).  Therefore the factorial divergence in
$\cR^{\PT}$ originating from the low-momentum integration region in
the Feynman diagrams, $k\sim Q e^{-\ell}$, is subtracted off exactly
by $\cR^{\merge}$.  The mismatch vanishes at large $\ell\gg
\ln\frac{Q}{\mu_I}$.  Due to this cancellation, the difference
$(\cR^{\PT} - \cR^{\merge})$ is well defined perturbatively, so we can
truncate the perturbative expansions for both $\cR^{\PT}$ and
$\as^{\PT}(k^2)$ in $\cR^{\merge}$ at some finite $N^{th}$ order.

The merging piece truncated at $\ell= N\!-\!1$ takes the form
\begin{equation}
   \cR^{\merge}(Q,\mu_I;N) = 
 \rho^{(V)}\>\frac{4C_F}{\pi^2}\>\cM\> \frac{\mu_I}{Q}
\>  \as
\sum_{\ell=0}^{N-1} \left(\be_0\frac{\as}{2\pi}\right)^\ell C_\ell\>.
\end{equation}
The final answer  reads 
\begin{equation}\label{finalmerge}
 \cR(Q) = \cR^{\PT}(Q;N) 
+  \rho^{(V)}\>\frac{4C_F}{\pi^2}\>\cM\> \frac{\mu_I}{Q}
\left(\alpha_0-  \as
\sum_{\ell=0}^{N-1} \left(\be_0\frac{\as}{2\pi}\right)^\ell C_\ell\right)
+\Delta_\cR(Q,\mu_I;N)\>.
\end{equation}
The answer contains two components, a perturbative piece with a smooth
logarithmic $Q$-dependence, $\cR^{\PT}=\cR^{\PT}(\as(Q^2))$, and a
steep power term $\cR{\NP}-\cR^{\merge}\propto \mu_I/Q$.
Correspondingly, the error $\Delta_\cR$ of the representation
\eqref{finalmerge} contains two separate components.

For a moderate $N$ the error in the perturbative part is
obviously given by
\begin{equation}\label{errPT}
 \Delta_\cR^{\PT} \>=\> \cO{\as^{N+1}(Q^2)}\>.
\end{equation}
The power-behaving component contains several sources. 
The first is the truncation error in $\cR^{\merge}$ 
which amounts to
\begin{equation}\label{errmerge}
 \Delta_\cR^{\merge} \>=\>
 \cO{\as(Q^2)\,\lambda^{N}\,\frac{\mu_I}{Q}}=\>
 \cR^{\NP} \cdot \cO{\frac{\as}{\al_0} \lambda^{N}}
 \>,\qquad 
  \lambda\equiv \beta_0\frac{\as}{2\pi}\ln\frac{Q}{\mu_I}\> < \> 1\>.
\end{equation}
The $\lambda$ parameter is smaller than unity provided that
$\mu_I>\lqcd$.  It is important to stress that these estimates are
uniform in $N$, since the factorially growing coefficients in
$\cR^{\PT}$ and $\cR^{\merge}$ cancel.  Constructing the difference of
the corresponding renormalon contributions, for the remainder of the
series for $\cR^{\PT}(Q,N)-\cR^{\merge}(Q,\mu_I;N)$ we have
$$
\as\left(\be_0\frac{\as}{2\pi}\right)^{N} \int_0^{\ln\frac{Q}{\mu_I}}
dt\, t^N\, e^{-t}  
\>\simeq\> \as(Q^2) \lambda^N \,\frac{\ln(Q/\mu_I)}{(N+1)}\,
 \frac{\mu_I}{Q}\>\ll\> \Delta_\cR^{\merge}
 , \qquad \mbox{for}\>\> N\gg\ln \frac{Q}{\mu_I}\>.
$$
The renormalon leftover being smaller than \eqref{errmerge} prevents
the magnitudes of the errors in \eqref{errPT} and \eqref{errmerge}
from exploding at large $N$.

Another error originates from the cutoff on the momentum integral 
in \eqref{A1cut} at the finite value $\mu_I$:
\begin{equation}\label{errNP}
 \Delta_\cR^{\NP} \>=\> \cO{\as^{\NP}(\mu_I^2)\frac{\mu_I}{Q}}
 = \>\cR^{\NP} \cdot \cO{\frac{\as^{\NP}(\mu_I^2)}{\al_0}}
 \>.
\end{equation}
Within the logic of the present approach, $\as^{\NP}$ is a rapidly
falling function, so that by choosing a sufficiently large $\mu_I$ one
can make this contribution arbitrarily small.

The largest contribution to the error in the power-behaved piece will
of course come from  higher corrections to the Milan factor which will 
have an effect of the form 
$$
\cM \>\Longrightarrow\> \left(1 + \cO{\frac{\as}{\pi}}\right)\,\cM\,,
$$
where $\as$ enters at some small scale. To quantify such effects
one has to go beyond a two-loop calculation. Whether they are
reasonably small depends on the effective interaction strength at
small scales. The present-day phenomenological estimate of
$\as/\pi\simeq0.2$ may lead one to be hopeful.

To conclude our discussion of the errors, we note that the difference
$(\cR^{\PT}\!-\! \cR^{\merge})$, the ``regularised'' perturbative
contribution to the observable, essentially corresponds to the
introduction of an infrared cutoff $\mu_I$ in the frequencies in
Feynman diagrams.  The regularised perturbative contribution, as
expected, depends on the infrared cutoff $\mu_I$.

The $\mu_I$-dependences in $\cR^{\NP}(\mu_I)$ and
$\cR^{\merge}(Q,\mu_I;N)$ compensate each other.  With increasing $N$,
the residual $\mu_I$-dependence disappears.  One can verify this by
evaluating the derivative of \eqref{finalmerge} with respect to
$\mu_I$:
$$
\frac{\partial}{\partial\ln\mu_I} \cR(Q) \>= \>
\cO{\Delta_\cR^{\merge}
+\Delta_\cR^{\NP} }. 
$$

A final remark concerns the scheme dependence.  Explicit expressions
for the series for both $\cR^{\PT}(Q,N)$ and $\cR^{\merge}(Q,\mu_I;N)$
depend on the scheme, that is on the choice of the perturbative
expansion parameter $\as$.  Throughout our analysis we have been using
the so-called physical (CMW) scheme~\citd{CMW}{DKT} in which the
intensity of soft gluon radiation equals, by definition, $\as$.
Shifting, for example, to the popular $\MSbar$ scheme amounts to
substituting 
\begin{equation}
  \label{Kdef}
 \as \>\Longrightarrow\> \alpha_{\MSbar} 
+ K \frac{\alpha_{\MSbar}^2}{2\pi} + \ldots 
\end{equation}
in the perturbative series, with $K$ as defined in \eqref{eq:Kdef}. 
As for the non-perturbative parameter $\al_0$ (and $\al'_0$, see below) 
it is {\em defined}\/ in \eqref{a0def} in terms of the CMW coupling.
Changing the scheme would change the {\em expression}\/ in the 
left-hand side of \eqref{a0def}, but would not affect the {\em value}\/
of $\al_0$. 
 
Notice that the normalisation of the second, power behaving, term 
in \eqref{finalmerge} namely,
$$
\cR^{\NP}(Q,\mu_I)-\cR^{\merge}(Q,\mu_I;N)\>,
$$ 
differs by the factor $2\cM/\pi$ from that of~\cite{DW} which is being
currently used in the experimental analyses. We return to this point
in the conclusions.

\subsection{$N=2$ merging}
For the practically important case $N=2$, when the perturbative answer
is known at the second order in the coupling constant $\as\equiv
\as(Q^2)$, the relevant moments determining the
power contributions to $T,D,C$ and $B$ radiators read 
\begin{equation}
\label{A1A11subtracted}
  A_1 \simeq  \frac{2C_F}{\pi^2} \mu_I
\left\{
\alpha_0(\mu_I)- 
\as
- 
\beta_0\frac{\as^2}{2\pi}\left(\ln\frac{Q}{\mu_I}+1\right)
\right\}\>,
\end{equation}
and
\begin{equation}\label{A11+A1}
 \half A_1'+A_1  \simeq  \frac{2C_F}{\pi^2} \mu_I
\left\{
\alpha_0'(\mu_I)+
\as 
+ 
\beta_0\frac{\as^2}{2\pi}\left(\ln\frac{Q}{\mu_I}+2\right)
\right\}\>.
\end{equation}
Here the non-perturbative parameter $\alpha_0$ has been defined in 
\eqref{a0def},
and $\alpha_0'$ is given by a similar integral 
but with an extra logarithmic factor: 
\begin{equation}\label{a0prime}
 \alpha_0(\mu_I) \equiv \int_0^{\mu_I} \frac{dk}{\mu_I}\>\as(k^2)\>,
\qquad
\alpha_0'(\mu_I) \equiv 
\int_0^{\mu_I} \frac{dk}{\mu_I}\>\as(k^2)\>\ln\frac{k}{\mu_I}\> .
\end{equation}
Let us note that from \eqref{a0prime} we expect the new
non-perturbative parameter $\alpha_0'$ to be {\em negative}.
Using the dispersive representation \eqref{asaedisp}, it is
straightforward to  derive
the following relation for $\half A_1'+A_1$
\begin{equation}
 \half A_1'+A_1 = \frac{2C_F}{\pi^2} \int_0^\infty dk\> 
 \alpha^{\NP}(k^2)\>\ln\frac{k}{\mu}\,.
\end{equation}
In analogy with the treatment of the $A_1$ moment, \eqref{A1cut}, we
truncate the integral on the right-hand side at the matching scale,
$\mu_I$: 
$$
 \frac{2C_F}{\pi^2} \int_0^\infty
 dk\>\alpha^{\NP}(k^2)\>\ln\frac{k}{\mu} 
\> \approx\> \frac{2C_F}{\pi^2} \int_0^{\mu_I} dk\>
              \alpha^{\NP}(k^2)\>\ln\frac{k}{\mu}\> .
$$

\paragraph{Linear observables.}
Thus, the merging contribution to $V=1\!-\!T,D,C$
distributions is, to second order, 
\begin{equation}
  \cR_{(V)}^{\merge}(Q,\mu_I;2) = 
\rho^{(V)}\cdot\frac{4C_F}{\pi^2}\>\cM\>
  \frac{\mu_I}{Q}
\left(\as + \be_0\frac{\as^2}{2\pi}
  \left(\ln\frac{Q}{\mu_I}+1\right)\right),
\end{equation}
and the final answer takes the form
\begin{equation}
  \cR= \cR^{\PT}(Q;2)+\rho^{(V)}\cdot\frac{4C_F}{\pi^2}\>\cM\>
 \frac{\mu_I}{Q}
  \left[\,\alpha_0 - \as -\be_0\frac{\as^2}{2\pi}
  \left(\ln\frac{Q}{\mu_I}+1\right)\,\right].
\end{equation}

\paragraph{Log-Linear observable.}
Substituting \eqref{A1A11subtracted} into the non-perturbative
broadening radiator \eqref{broadpower}, we arrive at
\begin{align}
  \cR_{(B)}^{\NP}
&= 2\nu\,\frac{4C_F}{\pi^2}\,\cM\> 
  \frac{\mu_I}{Q}\left[\,
  \alpha_0\left(\ln\frac{Q}{\mu_I}-\xi\right)
  -\alpha_0'\,\right]\>; \\
  \cR_{(B)}^{\merge}
&= 2\nu\,\frac{4C_F}{\pi^2}\,\cM\> 
  \frac{\mu_I}{Q}
\left[\, \as\left(\ln\frac{Q}{\mu_I}-\xi+1\right)
 +\be_0\frac{\as^2}{2\pi}
\left( \ln^2\frac{Q}{\mu_I}
      +(2-\xi)\left(\ln\frac{Q}{\mu_I}+1\right) \right) \,\right].
\end{align}
The final answer takes the form
\begin{equation}
\begin{split}
  \cR_{(B)}= \cR^{\PT} 
+ 2\nu\,\frac{4C_F}{\pi^2}\,\cM\> 
  \frac{\mu_I}{Q}\left[\, \left(\ln\frac{Q}{\mu_I}-\xi\right)
  \left(\alpha_0 - \as -\be_0\frac{\as^2}{2\pi}
  \left(\ln\frac{Q}{\mu_I}+1
  \right)\right)\right.
\\
  - \left. \left(\alpha_0' + \as +\be_0\frac{\as^2}{2\pi}
  \left(\ln\frac{Q}{\mu_I}+2 \right)\right) \,\right].
\end{split}
\end{equation}

\section{Summary and discussion}

In this paper we have considered the group of event shape observables
that exhibit $1/Q$ non-perturbative corrections. We have shown that
the two-loop effects result in a rescaling of the ``naive''
perturbative estimate of the magnitude of the power terms by a {\em
  universal}\/ Milan factor.  This is true for the non-perturbative
effects in the thrust, in the invariant jet mass and in the
$C$-parameter distributions, and for the main $\ln Q$--enhanced power
term in jet broadening.  The same Milan factor also applies to the
energy-energy correlation measure away from the back-to-back region.

The universality of the Milan factor for the linear jet shape
observables in DIS has recently been demonstrated by M.~Dasgupta and
B.R.~Webber in~\cite{DWDISMilan}.

We have shown, following \cite{DW97} that the $1/Q$ power effects in
jet shape distributions result in a {\em shift}\/ of the perturbative
spectra,
\begin{align}
\label{answ1} 
\frac{d\sigma}{ dV}(V) &=\left[\,\frac{d\sigma}{ dV}
\left(V-c_V\cP\right)\,\right]_{\PT} \qquad \qquad \qquad 
V=1\!-\!T,\> C, \>\frac{M_T^2}{Q^2},\> \frac{M^2_H}{Q^2}\>; \\
\label{answ2} 
 \frac{d\sigma}{ dV}(V) &=\left[\,\frac{d\sigma}{ dV}
\left(V-c_V\cP\ln \frac{Q}{Q_B}\right)\,\right]_{\PT} \qquad \>\> 
V=B_T, \>B_W\>.
\end{align}
The all-order resummed expressions for the perturbative spectra 
in \eqref{answ1} can be found, for the thrust and jet-masses, in
\cite{CTTW}, and for the $C$-parameter in \cite{CW}.
For the perturbative broadening spectrum \eqref{answ2}
one should use the recent theoretical prediction \cite{DLMSbroad},
which improves the treatment of subleading effects compared to the
original derivation \cite{CTW}.

The answers are valid in the region $1\gg V\gg \cP\sim \lqcd/Q$.
The relative coefficients $c_V$ are given in the table 
\begin{equation}
\mbox{
\begin{minipage}{4in}
\begin{tabular}{|l|c||c||c|c||c|c|} \hline 
 $V=$  & $ 1\!-\!T$ & $C$ & $M_T^2$ & $M^2_H$ & $B_T$ & $B_W$ \\ 
\hline
  $c_V=$  & 2 & $3\pi$ & 2 & 1 & 1 & $\half$ \\
\hline  
\end{tabular}
\end{minipage}
}
\end{equation}
Note that the power shifts in the single-jet
characteristics ($M^2_H, B_W$) amount to half of those in the
total distributions ($M^2_T, B_T$). 

The non-perturbative parameter $\cP$ was defined in \eqref{cPdef}. 
The $B$-distribution contains an additional non-perturbative
parameter, the scale of the logarithm $Q_B$, 
related to $\cP'$, \eqref{QBPprime},
the first $\log$-moment of the ``coupling modification'' $\de\ae$.
However, as was discussed in section~6, the parameters $\cP$ and $\cP'$
(or, $A_1$, $A'_1$) suffer from the same infrared renormalon ambiguity
as the perturbative series.
To remove this ambiguity one has to introduce the 
``infrared matching'' momentum scale $\mu_I$.  
At the two-loop level, the expressions for the shifts in terms of the
non-perturbative $\mu_I$-dependent phenomenological parameters 
$\al_0$~\eqref{a0def} and $\al'_0$~\eqref{a0prime} are: 
\begin{align}
\label{cPfin}
  \cP &\simeq \frac{4C_F}{\pi^2}\cM \frac{\mu_I}{Q}
\left\{
\alpha_0(\mu_I)- \as 
- \beta_0\frac{\as^2}{2\pi}\left(\ln\frac{Q}{\mu_I}+
\frac{K}{\be_0}+1\right)
\right\}\>, \\
\label{cPprfin}
 \cP'  &\simeq  \frac{4C_F}{\pi^2}\cM \frac{\mu_I}{Q}
\left\{\alpha_0'(\mu_I)+ \as 
+ \beta_0\frac{\as^2}{2\pi}\left(\ln\frac{Q}{\mu_I}+
\frac{K}{\be_0}+2\right)
\right\}\>; \quad \as\equiv \al_{\MSbar}(Q^2)\>.
\end{align}
Here we have given the $\MSbar$ expressions, with $K$ defined in
\eqref{eq:Kdef}.  
The parameter $\cP'$ enters into the shift in the broadening
distribution \eqref{answ2} as
\begin{equation}\label{QBtradedforcPprimewhynot?}
\cP\ln \frac{Q}{Q_B} \>=\> 
\cP\left(\ln \frac{Q}{\mu_I}-\xi\right)-\cP'\>,
\end{equation}
with the numerical parameter $\xi$ defined in \eqref{xidef}.  Recall
that we expect the parameter $\al_0'$ to be negative if the QCD
coupling is to remain positive in the infrared region, see
\eqref{a0prime}.  The $\ln Q$--enhancement of the $1/Q$ contribution
to jet broadening remains to be verified experimentally.

By now there is a good experimental evidence for $1/Q$ power terms in
jet shape variables both in $e^+e^-$~\cite{expee} and
DIS~\cite{expDIS}.  The magnitudes are consistent, within a 20\%\ 
margin, for different observables. An estimate of the basic $A_1$
parameter can be obtained, for example, from the second order fit to
the $Q$-dependence of the mean thrust,
\begin{equation}\label{A1est}
  \left\langle 1-T\right\rangle = a_1\as + a_2\as^2 +
   2\cP\>,
\end{equation}
with $a_1,a_2$ the known perturbative coefficients.  If one boldly
used the original relation \eqref{cPdef}, $\cP=\mbox{const}/Q$, the
fit to the data would produce $A_1\simeq 0.15\>\mbox{GeV}$.  However,
such an estimate is meaningless: the very representation \eqref{A1est}
would be renormalon-infested.  Therefore one should use instead in
\eqref{A1est} the representation \eqref{cPfin} for $\cP$ which is
constructed as the difference of the pure power contribution,
$\cR^{\NP}$, and the perturbative subtraction, $\cR^{\merge}$.  Thus,
the price one pays for avoiding the renormalon ambiguity is the
introduction of the finite infrared ``matching scale'' $\mu_I / \lqcd$
which enters into the shifts \eqref{cPfin}, \eqref{cPprfin} both
explicitly and via $\al_0$ and $\al_0'$.  As we have argued above in
section~6, the $\mu_I$-dependence gets weaker when higher orders of
the perturbative expansion are included.  The accuracy of the merging
of the perturbative and non-perturbative contributions is of order
$$
    \frac{\mu_I}{Q} 
     \left(\be_0\frac{\as(Q^2)}{2\pi}\ln\frac{Q}{\mu_I}\right)^{N+1} ,
$$
with $N$ the order of the highest perturbative term included
($N\!=\!2$ in \eqref{A1est}). A reasonable value to take for $\mu_I$
is $\mu_I=2\>\mbox{GeV}$.

Having considered the problem of the infrared renormalon divergence of
the perturbative series, we have chosen to quantify 
the non-perturbative effects in terms of the momentum integrals of the
running coupling $\as(k^2)$ rather than in terms of the moments of the
non-perturbative effective coupling modification, $\de\ae(m^2)$, as
was suggested in~\cite{DW}. 
Our non-perturbative parameter $\al_0$ 
differs by a factor $2/\pi$ from the analogous parameter 
$\bar{\al}_0$ of \cite{DW}. This factor originates from the
``translation'' relation  
$$
 \frac{2}{\pi}\> \int_0^\infty dk\>  \as^{\mbox{\scriptsize NP}}(k^2)
   \>=\>  \int_0^\infty d m\> \de\ae(m^2)\>.
$$
The power contributions $(\cR^{\NP}-\cR^{\merge})$ of \citd{DW}{DW97}
were written in terms of $\bar{\al}_0$ and did not include the
two-loop effects (the Milan factor).  Thus, to switch to the
expressions of this paper (based on $\al_0$, $\al'_0$), they should be
multiplied by $2/\pi\cdot\cM$.  These two factors practically
compensate each other: $2\cM/\pi\simeq 1.14$ ($n_f\!=\!3$).  Strictly
speaking, the remaining small renormalisation could affect the
phenomenological fits to the data, since it multiplies both the
non-perturbative term $\cR^{\NP}$ {\em and}\/ the perturbative
subtraction contribution ($\cR^{\merge}$), the latter possessing a
residual $\log Q$--dependence.

We could have chosen to express the answer in terms of the moments of
$\ae$ as suggested in \cite{DW}. This would have required, for the
sake of consistency, the reformulation also of the standard $\PT$
series and the $\PT$ subtraction in terms of $\ae$.  Carried out to
all orders, this would have given the same answer.  However, if we
restrict ourselves to $N=2$ we observe the following puzzle: neither
$\cR^{\PT}$ nor $\cR^{\merge}$ is sensitive to the difference between
$\as$ and $\ae$ to this order. As a result the coefficients of
$\cR^{\merge}(\ae)$ and $\cR^{\merge}(\as)$ differ by the factor
$2/\pi$, apparently for no good reason.  This mismatch can be
absorbed, but only partially, if one chooses an appropriate relation
between the cutoff scales $\mu_I$ in the two representations.  We
believe that the origin of the puzzle lies in the fact that the
relation between $\as$ and $\ae$ intrinsically contains the
non-perturbative region, and so only holds when one considers all
orders of the perturbation series.  We intend to consider this problem
in more detail in a future publication.

Concluding the discussion of the Milan factor we should stress that
only after having gone through the ordeal of the two-loop analysis,
can one unambiguously predict the relative magnitudes of the
power-behaving confinement contributions to different observables.
This is because the corrections to the magnitude of the power term
which are quadratic in small-scale $\as$, get promoted to a finite
renormalisation factor.  This can be understood with the help of the
following qualitative argument.  At the perturbative level, the
$\as^2$ correction can be recast in terms of a rescaling of $\lqcd$ in
the argument of $\as$ in the one-loop answer.  At the level of the
power terms, however, such a rescaling of the argument of the coupling
immediately triggers a corresponding rescaling of the dimensionful
{\em numerator}\/ of the $1/Q$ contribution.  In spite of the entirely
illustrative nature of the argument, it gives a hint as to why we do
not expect the higher loop corrections to be as crucially important as
the two-loop one. Indeed corrections beyond the second loop do not
affect the characteristic momentum scale of the problem and thus the
normalisation of the power contributions acquires only a factor of the
form $1+\cO{\as}$, which is formally subleading.  Whether the relative
size of such higher order corrections can be kept under control
depends on the actual strength of the effective interaction at small
scales and remains to be quantified.

Given the importance of the two-loop
effects, the ``naive'' first order approach becomes intrinsically
ambiguous.  Within this treatment one introduces into the Feynman
diagrams a coupling that runs with the gluon virtuality (fake gluon
mass) in order to define a ``projection'' onto the confinement
interaction region.  Such a procedure is non-unique with respect to
the precise determination of the argument of the coupling, and the
inclusion of the finite gluon mass effects into the kinematics and
into the definition of the observables.  For example, in the
definition of $T$ for the ($q\bar{q}+$ massive gluon) system suggested
by Beneke and Braun in~\cite{BB95}, the finite mass effects were
included in the normalisation of thrust ($p_q+p_{\bar{q}}+|\vec{k}_g|
< Q$).  Being as good a naive approximation as any other, this
produces an answer which differs by a numerical factor from the
``standard'' one.  Our claim is that if the two-loop effects were
evaluated, the final result based on the Beneke-Braun prescription
would coincide with the one obtained in \cite{DLMSthrust} and
generalised in the present paper.

Beyond the naive treatment, one can no longer embody all gluon decays
into the running of the coupling.  Those decay configurations that
affect the value of the observable under consideration cannot be
treated inclusively~\cite{NS}.  Though formally at the level $\as^2$
they are still essential, as we know, since they produce the same
leading $1/Q$ effects.  This is the Nason-Seymour problem of
non-inclusiveness of jet observables and it is a part of the whole
programme of giving unambiguous predictions for the magnitudes of
power-behaving contributions.  

Our solution can be formulated as follows.  First we give a definite
prescription for the naive contribution which treats gluon decays
inclusively and incorporates the running coupling and finite gluon
mass effects.  To this end we have chosen to substitute
$\sqrt{k_t^2+m^2}$ for $k_t$ both in the kinematics and in the
definition of the observable, so as to treat a ``massive gluon''
contribution.  Given this prescription, we have demonstrated that the
two-loop corrections to the naive treatment amount to a universal,
that is {\em observable-independent}, normalisation effect (Milan
factor).

The universality of the Milan factor has three ingredients.  Firstly,
it relies on universality of soft radiation, the latter being
responsible for confinement effects.  Secondly, it is based on the
concept of universality of the QCD interaction strength, all the way
down to small momentum scales, which has been processed through the
machinery of the dispersive approach.  Finally, it includes a certain 
geometric universality of the observables under consideration.  To
clarify the last point, we note that the contribution of a massless
parton $i$ to a given observable $V$ (linear in the limit of small
parton momenta) can be factorised as
$$
        v_i \>=\> k_{ti}\cdot f^{(V)}(\eta_i)\>,
$$
with $k_{ti}$ the modulus of transverse momentum and 
$\eta_i$ the rapidity of the parton, and $f$ a function that depends
on the observable. 
For example, for the thrust and $C$ parameter we have
\begin{equation*}
  f^{(T)}(\eta) = e^{-\eta}\Theta(\eta) + e^{\eta}\Theta(-\eta)  \>, 
  \qquad
  f^{(C)}(\eta) = \frac3{\cosh\eta} \>.
\end{equation*}
The rapidity integral of this function
enters linearly into the magnitude of the power contribution and
determines the observable-dependent coefficient of the latter,  
$$
\rho^{(V)} \>=\> \nu\cdot \int d\eta \> f^{(V)}(\eta)\>.
$$
The crucial observation is that this very same integral appears
both in the terms driven by a parent gluon (naive $\cR_0$ and the
inclusive correction $\cR^{in}$) as well as in the separate
contributions from offspring partons the gluon decays into (the
non-inclusive correction $\cR^{ni}$).  The two decay partons give
$$
v_1+v_2\>=\> k_{t1}f^{(V)}(\eta_1)+ k_{t2}f^{(V)}(\eta_2)\>,
$$
which contribution has to be weighted with the parent gluon decay
probability.  First we observe that the decay matrix element depends
only on the relative rapidity $\Delta\eta=\eta_1-\eta_2$ (Lorentz
invariance).  Then, that $\Delta\eta$ stays finite in the essential
region (IR-safety of $V$).  Therefore, keeping $\Delta\eta$ and
$k_{ti}$ fixed, we can perform an overall rapidity integration
extending it to infinity:
$$
v_1+v_2\Longrightarrow  \left(k_{t1}+k_{t2}\right)\cdot \rho^{(V)}\>.
$$
As a result, the dependence on the observable factors out thus
ensuring the universality of the non-inclusive correction. What is
left is the standard integral of the sum of two transverse momenta
weighted with the two-parton decay matrix element.

For some specific observables such as the height of the transverse
momentum distribution of Drell-Yan pairs at $q_t\!\to\!0$, $\EEC$ with
back-to-back kinematics, etc.\ we expect an interplay between the
perturbative resummation and the genuine confinement effects to lead
to calculable {\em fractional}\/ powers of $Q$ in the non-perturbative
correction terms.  Whether this expectation is correct is an
intriguing question which should be addressed both theoretically and
experimentally.  
The magnitude of these fractional power contributions
is determined by the non-per\-tur\-ba\-tive parameters $A_1$ and
$A_{2,1}$ (see \eqref{aefmom}).  The latter parameter enters also in the
$1/Q^2$ power corrections to DIS ($e^+e^-$) structure (fragmentation)
functions, DIS sum rules, the Drell-Yan $K$-factor, as well as in the
mean value of the 3-jet resolution parameter $y_{3}$~\citd{DMW}{DWA2}.
From the analysis of $F_2$, Dasgupta and Webber~\cite{DWA2} derived
a phenomenological estimate $A_{2,1}\simeq -0.2$~GeV$^2$.  
This implies that confinement effects
should {\em dampen}\/ the height of the Drell-Yan distribution at
$q_t=0$.

\vspace{2cm}
\noindent{\large\bf Acknowledgements}

\noindent
We are grateful to Martin Beneke, Vladimir Braun, Stefano Catani,
Mrinal Dasgupta, Gregory Korchemsky and Bryan Webber for illuminating
discussions and constructive criticism.

\end{document}